\documentstyle[12pt]{article}
\title{One-loop effective action for ${\cal N}=4$ SYM theory in the
hypermultiplet sector: leading low-energy approximation and beyond
}

\author{A.T. Banin\footnote{atb@math.nsc.ru}, I.L.
Buchbinder\footnote{joseph@tspu.edu.ru} , N.G.
Pletnev\footnote{pletnev@math.nsc.ru}}

\date{{\it
Institute of Mathematics, Novosibirsk, \\ 630090, Russia,\\
\vspace{0.7cm} Department of Theoretical Physics\\ Tomsk State
Pedagogical University\\ Tomsk 634041, Russia} }

\begin{document}

\begin{titlepage}
\maketitle \vspace{-10cm} \hfill{hep-th/0304046} \vspace{11cm}
\begin{abstract}
We develop the derivative expansion of the one-loop ${\cal N}=4$
SYM effective action depending both on ${\cal N}=2$ vector
multiplet and on hypermultiplet background fields.  Beginning with
formulation of ${\cal N}=4$ SYM theory in terms of ${\cal N}=1$
superfields, we construct the one-loop effective action with the
help of superfield functional determinants and calculate this
effective action in ${\cal N}=1$ superfield form using an
approximation of constant Abelian strength $F_{mn}$ and
corresponding constant hypermultiplet fields. Then we show that
the terms in the supercovariant derivative expansion of the
effective action can be rewritten in terms of ${\cal N}=2$
superfields. As a result, we get a new derivation of the complete
${\cal N}=4$ supersymmetric low-energy effective action obtained
in hep-th/0111062 and find subleading corrections to it. A problem
of ${\cal N}=4$ supersymmetry of the results is discussed.  Using
the formalism of ${\cal N}=2$ harmonic superspace and exploring
on-shell hidden ${\cal N}=2$ supersymmetry of ${\cal N}=4$ SYM
theory we construct the appropriate hypermultiplet-depending
contributions. The hidden ${\cal N}=2$ supersymmetry requirements
allow to get a leading, in hypermultiplet derivatives, part of the
correct ${\cal N}=4$ supersymmetric functional containing $F^{8}$
among the component fields.

\end{abstract}
\thispagestyle{empty}
\end{titlepage}

\newcommand{\be}{\begin{equation}}
\newcommand{\ee}{\end{equation}}
\newcommand{\bea}{\begin{eqnarray}}
\newcommand{\eea}{\end{eqnarray}}

\section{Introduction}
The ${\cal N}=4$ SYM theory attracts much attention due to the
remarkable properties allowing to clarify the profound questions
concerning the quantum dynamics in supersymmetric field models and
their links with string/brane theory. Maximally extended rigid
supersymmetry of the ${\cal N}=4$ SYM theory imposes strong
restrictions on the quantum dynamics. As a result, the quantities
characterizing the theory in quantum domain can be exactly found
or studied in great detail (see e.g. \cite{mald, bbiko, c, d, f}).

In this paper we calculate one-loop low-energy effective action in
${\cal N}=4$ SYM theory, depending on all fields of ${\cal N}=4$
vector multiplet. At present, the best, most symmetric and
adequate, description of ${\cal N}=4$ vector multiplet dynamics is
given in terms of unconstrained harmonic ${\cal N}=2$ superfields.
From this point of view, the ${\cal N}=4$ SYM theory is a model of
${\cal N}=2$ SYM theory coupled to hypermultiplet in adjoint
representation of a gauge group. It is well known that the exact
low-energy quantum dynamics of ${\cal N}=4$ SYM theory in ${\cal
N}=2$ vector multiplet sector is mastered by the non-holomorphic
effective potential ${\cal H}({\cal W},{\bar{\cal
W}})$\footnote{Low-energy effective action in arbitrary ${\cal N}=2$
SYM model can contain, in principle, a holomorphic effective potential
\cite{gates} but it vanishes in ${\cal N}=4$ gauge theory.}
, depending on ${\cal N}=2$ strengths ${\cal W}, {\bar{\cal W}}$ (see
Refs.  \cite{6, 9, grr, 7, bbku, bbiko}).  The explicit form of the
non-holomorphic potential for $SU(N)$ gauge group spontaneously broken
down to its maximal torus looks like
\begin{equation} {\cal H}({\cal
W}, \bar{\cal W}) = c\sum_{I<J}\, \ln \left({{\cal W}^{I}-{\cal
W}^{J}\over \Lambda}\right) \ln \left({\bar{\cal W}^{I}-\bar{\cal
W}^{J}\over \Lambda}\right),\label{1}
\end{equation}
where $\Lambda$ is
an arbitrary scale, $I,J = 1\ldots N$ and $c =1/(4{\pi})^2$ (for more
detail see Refs. \cite{bbku}).  Expression (\ref{1}) defines exact
low-energy effective potential in leading order in external momentum
expansion in ${\cal N}=2$ gauge superfield sector \cite{6, 9}. We
emphasize that the result (\ref{1}) is so general that it can be
obtained entirely on the symmetry grounds from the requirements of
scale independence and R-invariance up to a numerical factor \cite{6,
nonr}.  Moreover, the potential (\ref{1}) gets neither perturbative
quantum corrections beyond one-loop nor instanton corrections \cite{6,
9} (see also discussion of non-holomorphic potential in ${\cal N}=2$
SYM theories \cite{nonr, dwgr, zan, ins}).  All these properties are
very important for understanding of the low-energy quantum dynamics in
${\cal N}=4$ SYM theory in the Coulomb phase. In particular, the
effective potential (\ref{1}) provides the first subleading terms in
the interaction between parallel D3-branes in the superstring theory
(see e.g.  \cite{21}).  It was proposed that full ${\cal N}=4$ SYM
effective action, depending on proper invariants constructed from the
arbitrary powers of the Abelian strength $F_{mn}$ and obtained by
summing up all the loop quantum corrections, should reproduce (within
certain limits) the Born-Infeld action \cite{e} (${\cal N}=4$
SYM/supergravity correspondence). Discussion of this correspondence and
its two-loop test are given in Ref. \cite{21a} (see also a
consideration of the analogous problem for non-Abelian background in
Ref. \cite{Grasso} and general approach to calculating the higher loop
corrections in \cite{new}).

In order to clarify the structure of the restrictions on an
effective action, stipulated by ${\cal N}=4$ supersymmetry, and
to gain a deeper understanding of the ${\cal N}=4$
SYM/supergravity correspondence, we have to find an effective
action not only in ${\cal N}=2$ vector multiplet sector but
also depending on all the fields of ${\cal N}=4$ vector multiplet
(see discussion in \cite{iv}). This
problem remained unsettled for a long time. Recently, the
complete exact low-energy effective action containing the
dependence both on ${\cal N}=2$ gauge superfields and
hypermultiplets has been discovered \cite{31}. It has been shown that
the algebraic restrictions imposed by hidden ${\cal N}=2$
supersymmetry on a structure of the low-energy effective action in
${\cal N}=2$ harmonic superspace approach turn out to be so strong
that they allow to restore the dependence of the low-energy
effective action on the hypermultiplets on basis of the known
non-holomorphic effective potential (\ref{1}). As a result, the
additional hypermultiplet-dependent contributions containing the
on-shell ${\cal W}, {\bar{\cal W}}$ and the hypermultiplet
$q^{ia}$ \cite{33} superfields have been obtained in the form
\begin{equation}\label{2}
{\cal L}_{q}=c\left\{(X-1)\frac{\ln (1-X)}{X}+[{\rm
Li}_{2}(X)-1]\right\}, \quad X=-\frac{q^{ia}q_{ia}}{{\cal
W}\bar{\cal W}} ,
\end{equation}
where ${\rm Li}_{2}(X)$ is the Euler
dilogarithm function and $c$ is the same constant as in (\ref{1}) (see
the details and denotations in Refs. \cite{31, f}). The effective
Lagrangian (\ref{2}), together with the non-holomorphic effective
potential (\ref{1}), determine the exact ${\cal N}=4$
supersymmetric low-energy effective action in the theory under
consideration.

The leading low-energy effective Lagrangian (\ref{2}) has been
found in Ref. \cite{31} on a purely algebraic ground. It would
be extremely interesting to derive this Lagrangian and
next-to-leading corrections in external momenta in the framework
of the quantum field theory. Such a problem seems to be very
non-trivial since the expression (\ref{2}) includes any powers
of $X$ and is singular at ${\cal W}=0$, therefore the result can
not be obtained by considering the Feynman diagrams with the fixed
number of external hypermultiplet and gauge field legs. All such
diagrams must be summed up! In recent paper \cite{32}, the problem
of computing the effective Lagrangian (\ref{2}) has been solved
using the covariant harmonic supergraph techniques \cite{bbiko,
8}. The more general problem consists in quantum field theoretical
or algebraic derivation of the subleading terms in the effective
action, depending on all fields of ${\cal N}=4$ supermultiplet and
representation of these terms in complete ${\cal N}=4$ supersymmetric
form. The present paper is just devoted to solution of such a
problem for one-loop effective action. To be more precise, we
discuss the construction of the derivative expansion of one-loop
effective Lagrangian ${\cal L}_{eff}$ depending both on ${\cal
N}=2$ gauge background superfields, their spinor derivatives up to
some order and hypermultiplet background superfields using the
formulation of the ${\cal N}=4$ SYM theory in terms of ${\cal
N}=1$ superfields \cite{24, 27} and exploring the derivative
expansion techniques in ${\cal N}=1$ superspace \cite{15} (see
also \cite{ohr}). It allows us to obtain the exact coefficients at
various powers of covariant spinor derivatives of the ${\cal N}=2$
superfield Abelian strength ${\cal W}$ corresponding to the
constant space-time background that belongs to the Cartan subalgebra
of the gauge group $SU(N)$ spontaneously broken down to
$U(1)^{n-1}$ and constant space-time background hypermultiplet
$q^{ia}$
\begin{equation}\label{3}
{\cal W}|=\Phi={\rm const}, \quad D^{i}_{\alpha}{\cal
W}|=\lambda^{i}_{\alpha}={\rm const},\quad q^{ia}|={\rm const},
\end{equation}
$$
D^{i}_{(\alpha}D_{\beta )i}{\cal W}|=F_{\alpha\beta}={\rm
const},\quad D^{\alpha (i}D^{j)}_{\alpha}{\cal W}|=0,\quad
D_{\alpha}^i q^{aj}|=0, \quad D_{\dot{\alpha}}^i q^{aj}|=0,
$$
where $\Phi =diag(\Phi^{1},\Phi^{2},\ldots,\Phi^{n}),$
$\sum\Phi^{I}=0$. This background is the simplest one allowing
exact calculation of the one-loop effective action. We will show that
in this case the ${\cal N}=1$ superspace effective action can
be uniquely found on the basis of the effective action for vanishing
hypermultiplet \cite{bkts, 15} by means of a simple replacement of
variables. Following this, the obtained result is rewritten in a manifest
${\cal N}=2$ supersymmetric form using the same procedure as in
\cite{bkts} but maintaining the complete hypermultiplet dependence. We
emphasize that the background (\ref{3}) is a special
supersymmetric solution to classical equations of motion of the
${\cal N}=1$ superfield model representing the ${\cal N}=4$ SYM
theory in terms of ${\cal N}=1$ superfields and therefore the
effective action does not depend on the choice of ${\cal N}=1$
superfield gauge fixing conditions we impose on the theory.
Moreover, it can be shown that the background (\ref{3}) is completely
formulated in terms of ${\cal N}=2$ superfields, which provides a
possibility to write the effective action on this background in
a manifest ${\cal N}=2$ supersymmetric form. However, this
background is not form-invariant under the hidden ${\cal N}=2$
supersymmetry transformations of ${\cal N}=4$ supersymmetry. The
complete on-shell ${\cal N}=4$ supersymmetry involves the
transformations between the physical fields from the ${\cal N}=2$
vector multiplet and from hypermultiplets. But the background
(\ref{3}) does not contain the physical spinor fields from
hypermultiplets which are mixed with physical scalar fields from
${\cal N}=2$ vector multiplet under hidden ${\cal N}=2$
supersymmetry. Therefore it is likely from the outset that
the effective action on this background will not possess the
${\cal N}=4$ supersymmetry. The action is manifestly ${\cal N}=2$
supersymmetric but hidden extra ${\cal N}=2$ is violated. The only
term in the effective action that do not violate the hidden
supersymmetry is the one containing no spinor derivatives of ${\cal W}$
and hypermultiplets, which is just the effective potential
(\ref{2}). We will discuss these important points in Section 5.

The paper is organized as follows. In the next Section we recall
the known properties of ${\cal N}=4$ SYM theory in ${\cal N}=1$
and ${\cal N}=2$ formalism and discuss the background field
quantization including the choice of proper gauge fixing
conditions. In Section 3 we describe the calculations leading to
exact one-loop ${\cal N}=1$ superfield effective action for the
background (\ref{3}). Section 4 is devoted to representation of
this effective action in a manifest ${\cal N}=2$ form and discussing
the ambiguity of such a form. In Section 5 we demonstrate that
the first subleading term (containing the eighth power of Abelian
strength $F_{mn}$) in derivative expansion for the constructed
manifestly ${\cal N}=2$ supersymmetric effective action is
non-invariant under the hidden ${\cal N}=2$ supersymmetry
transformations of ${\cal N}=4$ SUSY and hence it is not ${\cal
N}=4$ supersymmetric.  Then we examine the hypermultiplet-dependent
terms which must be added to the known effective action in
the vector multiplet sector in order to make the whole $F^{8}$ term in
the complete effective action ${\cal N}=4$ supersymmetric. We show
that such terms can be found in the form of a finite order
polynomial in spinor derivatives of harmonic hypermultiplet
superfields and get the correct leading term in this polynomial.
In Summary we formulate final results and discuss unsolved
problems.

\section{Minimal formulation of ${\cal N}=4$ SYM theory in ${\cal N}=1,2$
superspaces and ${\cal N}=1$ supersymmetric background field method}

Formulation of ${\cal N}=4$ SYM theory possessing off-shell
manifest ${\cal N}=4$ supersymmetry is unknown so far. Therefore a
study of the concrete quantum aspects of this theory is usually
based on its formulation either in terms of physical component
fields (see e.g. \cite{Fr}) or in terms of ${\cal N}=1$ superspace
(see e.g. \cite{24}) or in terms of ${\cal N}=2$ harmonic
superspace \cite{gikos, 33}. In the first case, all four
supersymmetries are hidden; in the second case, one of them is
manifest and the other three are hidden; in the third case, two
supersymmetries are manifest and the other two are hidden. It is
worth pointing out that in all cases at least some of the
supersymmetries are on shell. Taking into account that the
presence of manifest symmetries simplifies a process of
calculations in quantum theory, it is reasonable to consider that
at present just ${\cal N}=2$ harmonic superspace formulation is
the best one for quantum ${\cal N}=4$ SYM theory. However, the
formulation in terms of ${\cal N}=1$ superspace has its own
positive features basically due to a relatively simple structure
of ${\cal N}=1$ superspace and large accumulated experience of
work with ${\cal N}=1$ supergraphs.

${\cal N}=4$ superfield description of ${\cal N}=4$ vector
multiplet can be realized with the help of on-shell ${\cal N}=4$
superfields $W^{AB}, A=1\ldots 4$ \cite{How} satisfying the
reality condition
$$
{W}^{AB}=\frac{1}{2}\varepsilon^{ABCD}W_{CD},\,
 W_{AB}=\bar{W}^{AB}
$$
and on-shell constrains
$$
 \bar{D}_{A\dot{\alpha}} W^{BC}=
\frac{1}{3}\delta_A^{[B}\bar{D}_{E\dot{\alpha}}W^{EC]},\,
D_{\alpha}^{(A} W^{B)C} =0.
$$
All physical fields of ${\cal N}=4$
vector multiplet are contained in the superfield $W^{AB}$. We
point out also the attempts to develop unconstrained formulation
in harmonic superspace approach \cite{Zup}. However, a ${\cal N}=4$
off-shell supersymmetric action for ${\cal N}=4$ SYM model is
still unknown.

\subsection{${\cal N}=4$ SYM theory in ${\cal N}=1$ superspace}
The physical field content of the superfield $W^{AB}$ can be
obtained by combining three ${\cal N}=1$ chiral superfields and
one ${\cal N}=1$ vector multiplet superfield \cite{24}. Then, the
six real scalars, which are the lowest components of the
superfield $W^{AB}$, are represented by the three complex scalar
components of the chiral ${\cal N}=1$ superfields ${\Phi}^{i}$.
The four Weyl fermions from $W^{AB}$ are divided into three plus
one. Three of them are considered as the spinor components of
${\Phi}^{i}$ and the fourth fermion is treated as gaugino and
constitutes, together with the real vector, the ${\cal N}=1$ vector
multiplet superfield $V$. In such a description, the
$SU(3)\bigotimes U(1)$ subgroup of $SU(4)$ $R-$ symmetry group is
manifest and the representations of the $SU(4)$ are decomposed
according to $\bf{6}\rightarrow \bf{3}+ \bf{\bar{3}}$,
$\bf{4}\rightarrow \bf{3}+\bf{1}$ so that the chiral superfields
$\Phi^{i}$ transform in the $\bf{3}$ of $SU(3)$, the antichiral
$\bar{\Phi}_{i}$ transforms in the $\bf{\bar{3}}$ and the vector
multiplet superfield is a singlet under $SU(3)$.

The action of ${\cal N}=4$ SYM model is formulated in terms of ${\cal N}=1$ superspace
as follows
\begin{eqnarray}
S &=& {1\over g^{2}}{\rm tr}\{\int d^{4}x
d^{2}\theta\,W^{2}+ \int d^{4}x
d^{4}\theta\,\bar{\Phi}_{i}{\rm e}^{V}\Phi^{i}{\rm
e}^{-V}+\nonumber\\
&+& {1\over 3!}\int d^{4}x d^{2}\theta\,
ic_{ijk}\Phi^{i}[\Phi^{j},\Phi^{k}]+{1\over 3!}\int
d^{4}x d^{2}\bar\theta\,
ic^{ijk}\bar\Phi_{i}[\bar\Phi_{j},\bar\Phi_{k}] \}.\label{classic}
\end{eqnarray}
The denotations and conventions correspond to Ref. \cite{24}.
All superfields here are taken
in the adjoint representation of the gauge group. Both ${\cal N}=1$
SYM and chiral superfield actions are superconformal invariants. In
addition to the manifest ${\cal N}=1$ supersymmetry and $SU(3)$
symmetry on the $i,j,k,\ldots$ indices of $\Phi$ and $\bar\Phi$, it has
hidden global supersymmetry given by transformations
\bea &\delta
W_{\alpha}=-\epsilon^{i}_{\alpha}\bar{\nabla}^{2}\bar\Phi_{c\,i}+
i\epsilon^{\dot\alpha}_{i}\nabla_{\alpha\dot\alpha}\Phi_{c}^{i},&\nonumber\\
&\delta \bar{W}_{\dot\alpha}=-\bar\epsilon_{\dot\alpha
i}\nabla^{2}\Phi_{c}^{i}+ i\epsilon^{\alpha
i}\nabla_{\alpha\dot\alpha}\bar\Phi_{c\,i}\label{hidden},&\\
&\delta \Phi_{c}^{i}=\epsilon^{\alpha i}W_{\alpha},\quad \delta
\bar\Phi_{c\,i}=\bar\epsilon^{\dot\alpha}_{i}\bar{W}_{\dot\alpha}.&\nonumber
\eea
The action (\ref{classic}) is also invariant under the transformations
 \bea \delta
\Phi_{c}^{i}&=&c^{ijk}\bar{\nabla}^{2}(\bar{\chi}_{j}\bar{\Phi}_{c\,k})+i
[{\chi}^{j}\bar\Phi_{c\,j},\Phi_{c}^{i}],\nonumber\\ \delta
\bar\Phi_{c\,i}&=&c_{ijk}\nabla^{2}({\chi}^{j}\Phi_{c}^{k})+i
[\bar{\chi}_{j}\Phi_{c}^{j},\bar\Phi_{c\,i}],\label{manifest}
\eea
Here the covariant spinor derivatives
${\nabla}_{\alpha}$,${\nabla}_{\dot{\alpha}}$,$ {\nabla}^{2}$ and
${\bar{\nabla}}^{2}$ are defined in Ref.  \cite{24} and
${\chi}^{i}$ are the ${\cal N}=1$ superfield parameters  forming
the $SU(3)$-isospinor as well as ${\Phi}^{i}$.  These parameters
include the central charge transformation parameters,
supersymmetry transformation parameters and internal symmetry
parameters of $SU(4)/SU(3)$. The transformations (\ref{manifest})
are given in terms of background covariant superfields $\Phi_{c} =
{\rm e}^{\bar\Omega}\Phi{\rm e}^{-\bar\Omega}$, $\bar{\Phi}_{c} =
{\rm e}^{-\Omega}\bar{\Phi}{\rm e}^{\Omega}$\, \cite{24}. Further
we use only these covariant chiral superfields and the
subscript $c$ is omitted.  It is convenient to introduce
the new notations
$\Phi^{1}=\Phi,\,\Phi^{2}=Q,\,\Phi^{3}=\tilde{Q}$ and rewrite two
last terms in (\ref{classic}) as follows
$$
i\int d^{4}x
d^{2}\theta\, Q[{\Phi},\tilde{Q}]+i\int d^{4}x d^{2}\bar\theta\,
\bar{Q}[\bar\Phi,\bar{\tilde{Q}}],
$$
which is ${\cal N}=1$ form
of hypermultiplet and lowest component of the chiral ${\cal N}=2$
field strength vector multiplet interaction for ${\cal N}=4$
model.

If the gauge group is Abelian, we get a free model. In the
non-Abelian case, the theory has a moduli space of vacua
parameterized by the vev's of the six real scalars. The manifold
of vacua is determined by the conditions of vanishing scalar
potential (F-flatness plus D-flatness) \cite{Arg}. The solutions
to the equations determining a vacuum structure of the theory can
be classified according to the phase of the gauge theory they give
rise to. In the pure Coulomb phase, each scalar field can have its
specific non-vanishing vev. As a result, space of vacua is ${\cal
M}=R^{6r}/{\cal S}_r$ where ${\cal S}_r$ is the Weyl group of
permutations for $r$ elements and an unbroken gauge group is $U(1)^r$.
But when several vev's coincide, some non-Abelian group $G \in
SU(N)$ remains unbroken and some massless gauge bosons appear in
the theory.

\subsection{${\cal N}=4$ SYM theory in ${\cal N}=2$ harmonic superspace}
From ${\cal N}=2$ supersymmetry point of view, the ${\cal N}=4$
vector multiplet consists of the ${\cal N}=2$ vector multiplet and
hypermultiplet. Therefore the ${\cal N}=4$ SYM action can be
treated as some special  ${\cal N}=2$ supersymmetric theory,
action of which is the action for ${\cal N}=2$ SYM theory plus
action describing the hypermultiplet $q^{ia}$ in adjoint
representation coupled to ${\cal N}=2$ vector multiplet. Such a
theory is formulated in ${\cal N}=2$ harmonic superspace
\cite{gikos, 33}. The dynamic variables in this case are real
unconstrained analytic gauge superfield $V^{++}$ and complex
unconstrained analytic superfield $q^{+}$. The harmonic gauge
connection $V^{++}$ serves as the potential of the ${\cal N}=2$
SYM theory and $q^{+}$ describes the hypermultipet. Action of the
${\cal N}=4$ SYM theory looks like
\begin{equation}
\label{act}
S[V^{++},q^{+},\breve{q}^{+}]=\frac{1}{2g^2}{\rm
tr}\int d^8 z \,{\cal W}^2 - \frac{1}{2g^2}{\rm tr}\int
d\zeta^{-4} q^{+a}{\cal D}^{++}q^{+}_{a}.
\end{equation}
The corresponding equations of motion are
\begin{equation}\label{eg}
D^{++}q^{+a} + ig\,[V^{++},q^{+a}]=0,\,
{\cal D}^{+\alpha}{\cal D}^+_{\alpha}{\cal W}= [q^{+a},q^+_a].
\end{equation}
Here $a=1,2$ is the index of the rigid $SU(2)$ symmetry,
$q^{+}_{a} =(q^{+},\breve{q}^{+})$,
$q^{+a}=\varepsilon^{ab}q^{+}_b =(\breve{q}^{+}, -q^{+})$ and
${\cal W}$ is the strength of ${\cal N}=2$ analytic gauge
superfield $V^{++}$ connection in the ${\lambda}$-frame
\cite{gikos, 33}, $g$ is a coupling constant, $d^8 z= d^4x
d^2\theta^+ d^2\theta^- du$, $d\zeta^{-4}= d^4x d^2\theta^+
d^2\bar{\theta}^+ du$, $du$ is the measure of integration over the
harmonic variables $u^{\pm i}$. The derivatives
$D^{+}_{\alpha(\dot{\alpha})}$ do not need a connection in the
frame where G-analyticity \cite{gikos, 33} is manifest. All other
denotations are given in Ref. \cite{33}. Equations (\ref{eg})
present the ${\cal N}=4$ SYM field equations of motion written in
terms of ${\cal N}=2$ superfields. The off-shell action
(\ref{act})  allows to develop the manifest ${\cal N}=2$
supersymmetric quantization. Moreover, this action is invariant
under hidden extra ${\cal N}=2$ supersymmetry transformations
\cite{33} which mix up $\cal W$, $\bar{\cal W}$ with $q^{+}_a$.
For our aim it is sufficient to point out that in the Abelian case
the corresponding transformations of hidden ${\cal N}=2$
supersymmetry are defined only on shell and have the form
\begin{eqnarray}
&\delta {\cal W}={1\over
2}\bar{\varepsilon}^{\dot{\alpha}\,a}\bar{D}^{-}_{\dot{\alpha}}q^{+}_{a},
\quad \delta \bar{\cal W}={1\over
2}{\varepsilon}^{\alpha\,a}D^{-}_{\alpha}q^{+}_{a}\label{hidden2},&
\\ &\delta q^{+}_{a}={1\over
4}(\varepsilon^{\alpha}_{a}D^{+}_{\alpha}{\cal W} +
\bar{\varepsilon}^{\dot\alpha}_{a}\bar{D}^{+}_{\dot\alpha}
\bar{\cal W}), \quad \delta q^{-}_{a}={1\over
4}(\varepsilon^{\alpha}_{a}D^{-}_{\alpha}{\cal W} +
\bar{\varepsilon}^{\dot\alpha}_{a}\bar{D}^{-}_{\dot\alpha}
\bar{\cal W}).& \nonumber
\end{eqnarray}
As a result, the model under consideration is ${\cal N}=4$
supersymmetric on shell.

Vacuum structure of the model (\ref{act}) in Abelian case is defined in
terms of solutions to
the following equations
\begin{equation}
\label{onsh} ({\cal
D}^+)^2{\cal W}=(\bar{\cal D}^+)^2\bar{\cal W} = 0,\, {D}^{++}q^{+a}=0,
\end{equation}
which are simple consequences
of the Eqs. (\ref{eg}) in Abelian case.
Equations (\ref{onsh}) for physical components of the ${\cal N}=4$
vector multiplet determined by the expansion
\begin{equation}\label{fizq}
q^+(\zeta,u)= f^i(x)u^+_i + \theta^{+\alpha}\psi_{\alpha}(x) +
\bar{\theta}^+_{\dot{\alpha}}\bar{\kappa}^{\dot\alpha}(x) +
2i\theta^+/\!\!\!\partial\bar{\theta}^+ f^i(x) u^-_i,
\end{equation}
$$ {\cal W}= \phi(x) + \theta^{-\,\alpha}\,\lambda_{\alpha}^{+}(x)\,+
\theta^{(+\alpha}\theta^{-\beta)}F_{\alpha\beta}(x),
$$
look like

\begin{equation}\label{free}
/\!\!\!\partial\psi =
/\!\!\!\partial\bar{\kappa} = \Box f^i = \Box \phi =
/\!\!\!\partial\lambda^i = \partial_{m}F_{mn}=0.
\end{equation}
The simplest solution to these equations of motion forms a set of
constant background fields
\begin{equation}\label{corset}
f^i={\rm const}, \psi ={\rm const}, \bar\kappa ={\rm const}, \phi
={\rm const}, F_{mn} ={\rm const}
\end{equation}
which transform {\it linearly} through each other under hidden
${\cal N}=2$ supersymmetry transformations (\ref{hidden2}):
\begin{eqnarray}
 &\delta\phi = \frac{1}{2}{\bar\varepsilon}^{\dot\alpha
a}{\bar\kappa}_{\dot\alpha a},\:
\delta\bar\phi = \frac{1}{2}\varepsilon^{\alpha a}\psi_{\alpha a},\:
\delta\psi_{\alpha a}= \frac{1}{2}\varepsilon^{\beta}_a
F_{\alpha\beta},\:\delta\bar\kappa_{\dot\alpha a}
=\frac{1}{2}\bar\varepsilon^{\dot\beta}_{a}
F_{\dot\alpha\dot\beta},& \nonumber\\
& \delta f^i_a = \frac{1}{4}\varepsilon^{\alpha}_a \lambda^i_{\alpha}+
\frac{1}{4}\bar\varepsilon_{a \dot\alpha}\bar\lambda^{i \dot\alpha},
\:\delta{\lambda}^i_{\alpha}= 0, \:
\delta F_{\alpha\beta}=0.&\label{delt}
\end{eqnarray}
The solution (\ref{corset}) is the simplest vacuum configuration
carrying out a representation of ${\cal N}=4$ supersymmetry and
allowing to calculate the ${\cal N}=4$ supersymmetric low-energy
effective action in ${\cal N}=4$ SYM theory.

It is instructive to compare the ${\cal N}=4$ supersymmetric
backgrounds (\ref{corset}) and the background (\ref{3}). Last
background contains the components $\phi$ and $F$ of the ${\cal
N}=2$ vector multiplet when the components $\bar\kappa$ and $\psi$
of the hypermultiplet are absent. As a result, the background
(\ref{3}) is not form-invariant under the hidden ${\cal N}=2$
supersymmetry transformations (\ref{delt}) and therefore this
background is not a representation of ${\cal N}=4$ supersymmetry.
However, the background (\ref{3}) is a representation of manifest
${\cal N}=2$ supersymmetry. Therefore we can state that effective
action found on the background (\ref{3}) within ${\cal N}=2$
background field method will be manifestly ${\cal N}=2$
supersymmetric and gauge invariant but it should not be completely
${\cal N}=4$ supersymmetric. The explicit calculations, fulfilled
in Sections 3 and 4, confirm this point of view. To construct the
complete ${\cal N}=4$ supersymmetric effective action, we can
follow the approach developed in \cite{31}. We consider the
effective action in ${\cal N}=2$ vector multiplet sector and find
such an action depending both on ${\cal N}=2$ vector multiplet and
on hypermultiplet so that the sum of above actions would be
invariant under hidden ${\cal N}=2$ supersymmetry transformations
(\ref{hidden2}). This idea is realized in Section 5 for finding
leading contributions to the ${\cal N}=4$ supersymmetric form of
the $F^{8}$-term in the effective action.

\subsection{${\cal N}=1$ background field quantization}
For computation of the effective action we use ${\cal N}=1$
superfield background field method (see e.g. \cite{24, 27})
in combination with ${\cal N}=1$ superfield heat kernel techniques
\cite{27}. These methods for constructing the effective action in
gauge field theories allow to preserve a classical gauge
invariance in quantum theory and sum, in principle, an infinite
set of Feynman diagrams to a single gauge invariant functional
depending on the background fields. As we pointed out, the theory
under consideration can be formulated either in terms of component
fields, or in terms of ${\cal N}=1$ superfields, or in terms of
${\cal N}=2$ harmonic superfields. The evaluation of effective
action in component formulation is extremely complicated even
within background field method because of a very large number of
interacting fields and the absence of manifest supersymmetry. The
effective action can be studied within ${\cal N}=2$ harmonic
superspace. The corresponding ${\cal N}=2$ background field method
was proposed in Refs \cite{8}. The aspects of heat kernel
techniques were considered in Refs. \cite{00}. However, these
techniques are undeveloped yet in many details to be extended for
our goals. We take into account a problem of matrix operators
mixing the ${\cal N}=2$ vector multiplet and hypermultiplet
sectors. Of course, the effective action can be studied using
harmonic supergraphs but we expect here meeting with the standard
problem of how to organize the calculations allowing to go beyond
leading low-energy approximation (see calculations in leading low-energy
approximation e.g. in Refs. \cite{bbku, 21a, 32, 8, 00}.
Therefore, we work within a formulation in terms of ${\cal N}=1$
superfields and use our experience to work with the theories
formulated in ${\cal N}=1$ superspace \cite{27, 15, new}.

The background field method is based on splitting the fields into
classical and quantum and imposing the gauge fixing conditions,
i.e. preservation of a classical gauge invariance,
only on quantum fields. Of course, the
gauge fixing conditions can break some classical symmetries as it
was mentioned above (for detail see e.g. Refs.
\cite{Holt, KuzMcTh}).
We define one-loop  effective action $\Gamma$ depending on the
background superfields (\ref{3}) by a path integral over quantum fields
in the standard form
\begin{equation}\label{4}
{\rm e}^{i\Gamma} = \int \,{\cal D}v\,\,{\cal D}\varphi\, {\cal
D}c\,{\cal D}c'\, {\cal D}\bar{c}\,{\cal D}\bar{c}'\, {\rm
e}^{i(S_{(2)}+S_{\rm FP})},
\end{equation}
where $S_{(2)}$ is a quadratic in quantum fields part of the
classical action including a gauge-fixing condition and $S_{\rm
FP}$ is a corresponding ghost action. Formal calculating the above
path integral leads to a functional determinant representation of
effective action (see (\ref{egamma})). The main technical tool we
use in this paper for the ${\cal N}=1$-superfield calculations is
the background covariant gauge-fixing multi-parametrical
conditions
\begin{equation}\label{fixing}
S_{\rm GF}=-{1\over \alpha g^{2}}\int
d^8z\,(F^{A}\bar{
F}^{A}+b^{A}\bar{b}^{A}),
\end{equation}
here $b, \bar{b}$ are the Nielsen-Kallosh ghosts. We choose the
convenient gauge-fixing conditions for the quantum superfields
$v$ and ${\varphi}$ in the form
\begin{eqnarray}
&\bar{F}^{A}=\nabla^2v^{A} +\lambda [{1\over
\Box_{+}}\nabla^{2}\varphi^{i},\bar{\Phi}_{i}]^{A}, & \nonumber\\
&F^{A}=\bar{\nabla}^2v^{A} -\bar{\lambda}[ {1\over
\Box_{-}}\bar{\nabla}^2\bar{\varphi}_{i},\Phi^{i}]^{A},&\label{f}
\end{eqnarray}
where $\alpha, \lambda, \bar{\lambda}$ are the arbitrary numerical
parameters and $\Box_+$, $\Box_-$ are the standard notations for
Laplace-like operators in the ${\cal N}=1$ superspace. It is
evident that the gauge fixing functions (\ref{f}) are covariant
under background gauge transformations. These gauge fixing
functions (\ref{f}) can be considered as a superfield form of
so-called $R_{\xi}$-gauges (see Refs. \cite{28, 23}) which are
usually used in spontaneously broken gauge theories. Since an
Abelian background is a solution to classical equations of motion,
we won't worry about the choice of gauge-fixing parameters.
Therefore it is convenient to take gauge-fixing which we
call the Fermi-DeWitt gauge: $\alpha=\lambda=1$. Such a choice of
the gauge parameters allows us to avoid the known problem
\cite{12} in the functional determinant method for calculating the
mixed contribution which contains vector-chiral superfield
propagators circulating along the loop. Calculation of such
contributions using a conventional gauge is related to the necessity
of working with a cumbersome expression:
\bea & {\rm Tr\,ln}(-\Box+
iW^{\alpha}\nabla_{\alpha}
+i\bar{W}^{\dot{\alpha}}\bar{\nabla}_{\dot{\alpha}} + M -&
\nonumber\\ &-\bar{X}\frac{1}{\Box_{-}-\mu\bar{\mu}}X\bar{
\nabla}^2{\nabla}^2
-X\frac{1}{\Box_{+}-\bar{\mu}\mu}\bar{X}\nabla^2\bar{\nabla}^2+
\bar{X}\frac{1}{\Box_{-}-\bar{\mu}\mu}\bar{\mu}\bar{X}
\bar{\nabla}^2+ X\frac{1}{\Box_{+}-\mu\bar{\mu}}\mu X
{\nabla}^2),&
\eea
where the notations of \cite{12} were used.

We want to note once again that in gauge theories not all rigid
symmetries of the classical action can be maintained manifestly in
quantum theory, even in the absence of anomalies. The issue here
is that quantization requires gauge fixing and the latter, as can
be shown, breaks some symmetries (breaking the classic conformal symmetry
is discussed e.g. in Ref. \cite{KuzMcTh}). This is the known
general situation (see e.g. \cite{Holt}). In our case, the
gauge-fixing (\ref{fixing}) obviously also breaks rigid classic ${\cal
N}=4$ symmetry (\ref{hidden}, \ref{manifest}) since it is
covariant only under ${\cal N}=1$ supersymmetry transformations.
Therefore the effective action obtained should be invariant only
under quantum deformed hidden transformations (\ref{hidden}). This
deformations can, in principle, be computed at each loop order but
this problem is beyond the purposes of this work.

After splitting each field
on the background and quantum part (i.e. ${\rm e}^{V_{tot}}={\rm
e}^{\Omega}{\rm e}^{g\,v}{\rm e}^{\bar\Omega}$,$\Phi\rightarrow \Phi
+ \varphi,\,\bar{\Phi}\rightarrow \bar{\Phi} + \bar{\varphi},\,
Q\rightarrow Q + q,\,\tilde{Q}\rightarrow \tilde{Q} + \tilde{q},\,
\bar{Q}\rightarrow \bar{Q} + \bar{q},\, \bar{\tilde{Q}}\rightarrow
\bar{\tilde{Q}} + \bar{\tilde{q}}$), we can rewrite a quadratic part
of sum of the classical action (\ref{classic}) and gauge fixing action
(\ref{fixing}) in the form
\begin{equation}\label{s2}
S_{(2)}= -{1\over 2}\sum_{I<J}\int d^{4}x d^{4}\theta\, ({\cal
F}^{IJ}{\bf H}_{IJ}{\cal F^{\dag}}^{IJ}+ \bar{v}^{IJ}(O_{V}-
M)_{IJ} v^{IJ}),
\end{equation}
 where $ {\cal F}= (\bar{\varphi},
{\varphi}, \bar{q}, q, \bar{\tilde{q}}, \tilde{q}), \quad {\cal F
}^{\dag}=({\varphi}, \bar{\varphi}, q, \bar{q}, \tilde{q},
\bar{\tilde{q}} )^{T},$
\begin{equation}\label{mass}
M_{IJ}=
(\bar{\Phi}_{IJ}\Phi_{IJ}+\bar{Q}_{IJ}Q_{IJ}+\bar{\tilde{Q}}_{IJ}\tilde{Q}_{IJ}),
\end{equation}
$$
O_{V}=\Box-iW^{\alpha}_{IJ}\nabla_{\alpha}-i\bar{W}^{\dot{\alpha}}_{IJ}\bar{\nabla}_{\dot{\alpha}},
$$
and $W^{\alpha}_{IJ}= W^{\alpha}_{I}-W^{\alpha}_{J},$ $
\bar{W}^{\dot{\alpha}}_{IJ}=\bar{W}^{\dot{\alpha}}_{I}-
\bar{W}^{\dot{\alpha}}_{J}$
are the background field strengths belonging to the Cartan subalgebra
and $ \Phi_{IJ}= \Phi_{I}-\Phi_{J}$. The Weyl basis in the space of
hermitian traceless matrices from the algebra $su(N)$ was used in
order to obtain (\ref{s2}). We consider the case of the gauge
group $SU(N)$ broken down to the maximal torus $U(1)^{N-1}$.
The constraint  $I<J$ arises since the components of the quantum
superfields which lie in the Cartan subalgebra do not interact
with the background field and therefore they completely decoupled.
For details of using the Weyl basis to calculate the effective action
see e.g. \cite{bbku}.

The operator ${\bf H}$ is a matrix depending on covariant derivatives
and background fields. The explicit form of this matrix looks like
\begin{equation}\label{matr}
\pmatrix{
G_{+}(\phi)\nabla^{2}\bar{\nabla}^{2}& 0 &
-\phi\bar{f}{\nabla^{2}\bar{\nabla}^{2}\over\Box}&
i\bar{v}\nabla^{2}& -\phi f{\nabla^{2}\bar{\nabla}^{2}\over\Box}&
-i\bar{f}\nabla^{2}\cr
0 & G_{-}(\phi)\bar{\nabla}^{2}\nabla^{2}&
iv\bar{\nabla}^{2} &
\bar{\phi}f{\bar{\nabla}^{2}\nabla^{2}\over\Box}&
-if\bar{\nabla}^{2}&
-\bar{\phi}v{\bar{\nabla}^{2}\nabla^{2}\over\Box}\cr
-f\bar{\phi}{\nabla^{2}\bar{\nabla}^{2}\over\Box}&
i\bar{v}\nabla^{2}&
G_{+}(f)\nabla^{2}\bar{\nabla}^{2}&
0 & f \bar{v}{\nabla^{2}\bar{\nabla}^{2}\over\Box}&
i\bar{\phi}\nabla^{2}\cr
-iv\bar{\nabla}^{2}&-\bar{f}\phi{\bar{\nabla}^{2}\nabla^{2}\over\Box}&
0 & G_{-}(f)\bar{\nabla}^{2}\nabla^{2}&
i\phi\bar{\nabla}^{2}&-\bar{f}v{\bar{\nabla}^{2}\nabla^{2}\over\Box}\cr
-v\bar{\phi}{\nabla^{2}\bar{\nabla}^{2}\over\Box}&
\bar{f}\nabla^{2}&-v\bar{f}{\nabla^{2}\bar{\nabla}^{2}\over\Box}&
-i\bar{\phi}\nabla^{2}&
G_{+}(v)\nabla^{2}\bar{\nabla}^{2}& 0 \cr
if\bar{\nabla}^{2}&-\bar{v}\phi{\bar{\nabla}^{2}\nabla^{2}\over\Box}&
-i\phi\bar{\nabla}^{2}&
-\bar{v}f{\bar{\nabla}^{2}\nabla^{2}\over\Box}&
0 & G_{-}(v)\bar{\nabla}^{2}\nabla^{2}
},
\end{equation}
where the following notations were used
$$
G_\pm(a)=1-{(a\bar{a})\over \Box_\pm}, \quad\phi=\Phi_{IJ}, \quad
\bar\phi=\bar\Phi_{IJ}, \quad f=Q_{IJ}, \quad
\bar{f}=\bar{Q}_{IJ}, \quad v=\tilde{Q}_{IJ}, \quad
\bar{v}=\bar{\tilde{Q}}_{IJ}
$$
and $\Box_\pm$ means
$\nabla^{2}\bar{\nabla}^{2}$ and $\bar{\nabla}^{2}\nabla^{2}$
respectively. In the space of chiral and antichiral superfields
these operators act as follows
$$
\nabla^{2}\bar{\nabla}^{2}=
\Box_{+}= \Box-i\bar{W}^{\dot{\alpha}}\bar{\nabla}_{\dot{\alpha}}-
{i\over 2}(\bar{\nabla}\bar{W}), $$ $$ \quad
\bar{\nabla}^{2}\nabla^{2}= \Box_{-}= \Box-i
W^{\alpha}\nabla_{\alpha}-{i\over 2}(\nabla W).
$$
It should be
noted that generally speaking, the second variation of the
classical action leads to a $7\times7$ matrix operator.
But the chosen gauge-fixing condition (\ref{f}) allows partial
diagonalization to a $1\times1\oplus 6\times6$ block matrix and
separation of kinetic operator for the vector fields. Hence
we avoid the functional trace calculation problem for the
above-mentioned cumbersome operator. This gauge-fixing condition
eliminates the interaction vertexes between quantum matter fields and
quantum vector fields but generates new interaction vertexes
between quantum chiral fields and ghosts.

Let us consider now a structure of Faddeev-Popov ghost contribution to
the one-loop effective action.
The action of the
Faddeev-Popov ghosts $S_{\rm FP}$ for the gauge fixing functions
(\ref{f}) has the form
\begin{equation}
S_{\rm FP}= {\rm tr}\int d^8z
\,\left( (\bar{c}'c-c'\bar{c})- \left( c' [\Phi^{i},{\lambda\over
\Box_{+}} [\bar{c},\bar{\Phi}_{i} ] ]+ \bar{c}' [{\bar\lambda\over
\Box_{-}} [c,\Phi^{i} ],\bar{\Phi}_{i} ] \right)\right).\label{acfp}
\end{equation}
It leads to the following contribution of the ghosts to the effective action
\begin{equation}\label{fp_matrix}
\ln ({\rm Det}({\bf H}_{FP})) = 2\sum_{I<J}{\rm Tr}\ln
\pmatrix{
0&(1-{M\over \Box_{+}})\nabla^{2}\bar{\nabla}^{2}\cr
-(1-{M\over \Box_{-}})\bar{\nabla}^{2}\nabla^{2}&0
}_{IJ},
\end{equation}
where $M$ was defined in (\ref{mass}).

The final result of the integration in path integral
(\ref{4}) over all quantum superfields is given by
formal representation for the one-loop effective action
in terms of functional determinants
\begin{equation}\label{egamma}
{\rm e}^{i\Gamma}=\prod_{I<J}{\rm Det}^{-1} (O_{V}-M){\rm Det}^{-1}
({\bf H}){\rm Det}^{2} ({\bf H}_{FP}).
\end{equation}
Since the strengths $\Phi$ and $W_{\alpha}$ belong to the Cartan
subalgebra, only half of roots should be taken into account during
the integration over the quantum fields and the effective action
looks like
$$
\Gamma = \sum_{I<J}\, \Gamma_{IJ}.
$$
Our next purpose is a computation of the
above functional determinants.

\section{Evaluations of superfield functional traces and one-loop effective action.}
In this section we present the basic steps of
functional traces calculations for the operators, which make
background-dependent contributions into the effective action
(\ref{egamma}). It is seen from (\ref{matr}) that in the absence of
background superfields  $Q, \tilde{Q}$, the matrix operator ${\bf H}$
includes only the background-dependent inverse propagators $G_+$, $G_-$
and vertices for the background field $\Phi$ interacting with the
quantum hypermultiplet. Such a situation has been studied
in detail (see e.g. Refs.
\cite{15, ohr, 12, bky}). It should be noted
that the form of ${\bf H}$ containing dressed inverse propagators
is directly related to the $R_{\xi}$ gauge fixing conditions
(\ref{f}).

On the first stage we divide the matrix ${\bf H}$ into a sum of two
matrices ${\bf H}={\bf H}_{\Box}+{\bf H}_{\nabla}$ where the matrix
(${\bf H}_{\Box}$) contains all blocks with
$\nabla^{2}\bar{\nabla}^{2},
\bar{\nabla}^{2}\nabla^{2}$ and the matrix ${\bf H}_{\nabla}$ contains blocks
with $\bar{\nabla}^{2}$ and $\nabla^{2}$ only. Let's present the logarithm
of matrix $\ln ({\bf H}_{\Box})$ as follows
$$
\ln ({\bf H})= \ln ({\bf
H}_{\Box})+\ln (1- ({\bf H}_{\Box}^{-1}{\bf H}_{\nabla})).
$$
Using the known Frobenius formula for inversion
of block type matrix
$$
H = \pmatrix{A & B\cr C & D}, \quad
H^{-1}= \pmatrix{A^{-1}+A^{-1}BE^{-1}CA^{-1} & -A^{-1}BE^{-1}\cr
-E^{-1}CA^{-1} & E^{-1}},
$$
where $E = D - CA^{-1}B$,
we get by direct calculation the inverse matrix for ${\bf H}_{\Box}$
\begin{equation}\label{matrinv}
{\bf H}^{-1}_{\Box}= \pmatrix{
g_{+}(\phi){\nabla^{2}\bar{\nabla}^{2}\over\Box_{+}^{2}}& 0 &
{\phi\bar{f}\over\Box_{+M}}{\nabla^{2}\bar{\nabla}^{2}\over\Box_{+}^{2}}
& 0 &
{\phi\bar{v}\over\Box_{+M}}{\nabla^{2}\bar{\nabla}^{2}\over\Box_{+}^{2}}
& 0 \cr 0 &
g_{-}(\phi){\bar{\nabla}^{2}\nabla^{2}\over\Box_{-}^{2}}& 0 &
{\bar{\phi}f\over\Box_{-M}}{\bar{\nabla}^{2}\nabla^{2}\over\Box_{-}^{2}}
& 0 &
{\bar{\phi}v\over\Box_{-M}}{\bar{\nabla}^{2}\nabla^{2}\over\Box_{-}^{2}}
\cr
{f\bar{\phi}\over\Box_{+M}}{\nabla^{2}\bar{\nabla}^{2}\over\Box_{+}^{2}}
& 0 & g_{+}(f){\nabla^{2}\bar{\nabla}^{2}\over\Box_{+}^{2}} & 0 &
{f
\bar{v}\over\Box_{+M}}{\nabla^{2}\bar{\nabla}^{2}\over\Box_{+}^{2}}
& 0 \cr 0 &
{\bar{f}\phi\over\Box_{-M}}{\bar{\nabla}^{2}\nabla^{2}\over\Box_{-}^{2}}
& 0 & g_{-}(f){\bar{\nabla}^{2}\nabla^{2}\over\Box_{-}^{2}} & 0 &
{\bar{f}v\over\Box_{-M}}{\bar{\nabla}^{2}\nabla^{2}\over\Box_{-}^{2}}\cr
{v\bar{\phi}\over\Box_{+M}}{\nabla^{2}\bar{\nabla}^{2}\over\Box_{+}^{2}}&
0 &
{v\bar{f}\over\Box_{+M}}{\nabla^{2}\bar{\nabla}^{2}\over\Box_{+}^{2}}&
0 & g_{+}(v){\nabla^{2}\bar{\nabla}^{2}\over\Box_{+}^{2}}& 0 \cr 0
&{\bar{v}\phi\over\Box_{-M}}{\bar{\nabla}^{2}\nabla^{2}\over\Box_{-}^{2}}&
0 &
{\bar{v}f\over\Box_{-M}}{\bar{\nabla}^{2}\nabla^{2}\over\Box_{-}^{2}}&
0 & g_{-}(v){\bar{\nabla}^{2}\nabla^{2}\over\Box_{-}^{2}}  },
\end{equation}
here we have introduced the notations
$$
g_\pm({\phi})=1+{{{\phi}{\bar{\phi}}}\over{\Box_{\pm M}}},
{\Box_{\pm M}}={\Box}_{\pm} -{M}.
$$
One can note that the
combination $M$ (\ref{mass}) has naturally appeared during the
inversion procedure. Then we find the product ${\bf
H}_{\Box}^{-1}{\bf H}_{\nabla}$ in a remarkably simple form
\begin{equation}
{\bf H}_{\Box}^{-1}{\bf H}_{\nabla}= \pmatrix{ 0 & 0 & 0 & i\bar{v}
{\nabla^{2}\over \Box_{-}} & 0 & - i\bar{f}{\nabla^{2}\over
\Box_{-}}\cr 0 & 0 & iv{\bar\nabla^{2}\over \Box_{+}} & 0 & -
if{\bar\nabla^{2}\over \Box_{+}}& 0 \cr 0 &-i\bar{v}{\nabla^{2}\over
\Box_{-}} & 0 & 0 & 0 &i\bar{\phi}{\nabla^{2}\over \Box_{-}}\cr
-iv{\bar\nabla^{2}\over \Box_{+}} & 0 & 0 & 0 & i\phi
{\bar\nabla^{2}\over \Box_{+}}& 0 \cr 0 & i\bar{f}{\nabla^{2}\over
\Box_{-}} & 0 & -i\bar{\phi}{\nabla^{2}\over \Box_{-}}& 0 & 0 \cr
if{\bar\nabla^{2}\over \Box_{+}}& 0 &-i\phi{\bar\nabla^{2}\over
\Box_{+}}& 0 & 0 & 0 }
\end{equation}

The next stage consists in matrix traces calculations. Let's
expand ${\rm Tr}(\ln (1- ({\bf H}_{\Box}^{-1}{\bf H}_{\nabla})))$
in a series in powers of ${\bf H}_{\Box}^{-1}{\bf H}_{\nabla}$.
The nonzero matrix traces will have only even powers of the
series, which are grouped into
\begin{equation}\label{h1}
{\rm Tr}_{6\times 6}(\ln (1- \ln ({\bf H}_{\Box}^{-1}{\bf H}_{\nabla})))=
{\rm Tr}\left(\ln \left(1-{M\over
\Box_{+}}\right){\nabla^{2}\bar{\nabla}^{2}\over\Box_{+}}\right)+
{\rm Tr}\left(\ln \left(1-{M\over
\Box_{-}}\right){\bar{\nabla}^{2}\nabla^{2}\over\Box_{-}}\right),
\end{equation}
where $M$ was introduced in ({\ref{mass}) and ${\rm Tr}$ means the
functional trace. Also we have to consider the matrix trace of the
$\ln ({\bf H}_{\Box})$. According to the above strategy,
we write the matrix as a diagonal matrix plus the rest, i.e. ${\bf
H}_{\Box}={\bf H}_{0}+{\bf \Delta}$:
\begin{equation}\label{split_m}
{\rm Tr}\ln ({\bf H}_{\Box})= {\rm Tr}\ln ({\bf H}_{0})+ {\rm Tr}\ln (1+{\bf H}_{\Box}^{-1}{\bf
\Delta}),
\end{equation}
where matrix ${\bf H}_{0}$ contains only $\nabla^{2}\bar\nabla^{2}$
and $\bar\nabla^{2}\nabla^{2}$ at zero background fields $\Phi, Q,
\tilde{Q}$ and
therefore can be omitted. The matrix elements of ${\bf H}_{\Box}^{-1}{\bf \Delta}$ are
blocks with chiral ${\nabla^{2}\bar{\nabla}^{2}\over\Box_{+}}$ and antichiral
${\bar{\nabla}^{2}\nabla^{2}\over\Box_{-}}$ projectors. After permutation of the
lines and columns, the trace logarithm of the matrix $1+{\bf H}_{0}^{-1}{\bf
\Delta}$ can be reorganized as follows
\begin{equation}
{\rm Tr}_{6\times 6}\ln (1+{\bf H}_{0}^{-1}{\bf \Delta}) = {\rm Tr}_{3\times 3}\ln \left(
1-
\pmatrix{
(\phi\bar\phi)&(\phi\bar{f})&(\phi\bar{v})\cr
(f\bar\phi)&(f\bar{f})&(f\bar{v})\cr
(v\bar\phi)&(v\bar{f})&(v\bar{v})
}
{\nabla^{2}\bar{\nabla}^{2}\over \Box_{+}^{2}}
\right)+ ({\nabla^{2}\bar{\nabla}^{2}\over \Box_{+}^{2}}\rightarrow
{\bar{\nabla}^{2}\nabla^{2}\over \Box_{-}^{2}}).
\end{equation}
The direct calculation of the matrix traces for the first terms in the Taylor series
allows to write the result
\begin{equation}\label{1-dm}
{\rm Tr}_{6 \times 6}\ln (1+{\bf H}_{0}^{-1}{\bf \Delta})=
{\rm Tr}\left(\ln (1-{M\over \Box_{+}}){\nabla^{2}\bar\nabla^{2}\over
\Box_{+}}\right)+
{\rm Tr}\left(\ln (1-{M\over \Box_{-}}){\bar\nabla^{2}\nabla^{2}\over
\Box_{-}}\right).
\end{equation}
that together with (\ref{h1}) gives
\begin{equation}\label{hresult}
\ln ({\rm Det}^{-1}({\bf H}))=-2{\rm Tr}\left(\ln
(1-{M\over \Box_{+}}){\nabla^{2}\bar\nabla^{2}\over
\Box_{+}}\right)- 2{\rm Tr}\left(\ln (1-{M\over
\Box_{-}}){\bar\nabla^{2}\nabla^{2}\over
\Box_{-}}\right).
\end{equation}

The contribution of the Faddeev-Popov ghosts is determined by
eq.(\ref{fp_matrix}). Extracting and neglecting the expression
$\ln \left(\pmatrix{0 & \nabla^2\bar{\nabla}^2\cr -\bar{\nabla}^2\nabla^2 &
0}\right)
$, we obtain the ghost contribution to the effective action in the form
\begin{equation}\label{fpresult}
\ln ({\rm Det}^2({\bf H}_{FP}))= 2{\rm Tr}\left(\ln (1-{M\over \Box_{-}}){\bar{\nabla}^{2}\nabla^{2}\over
\Box_{-}}\right)
 + 2{\rm Tr}\left(\ln(1-{M\over \Box_{+}}){\nabla^{2}\bar{\nabla}^{2}\over \Box_{+}}\right),
\end{equation}
which is exactly (\ref{hresult}) with the opposite sign. Therefore
second and third functional determinants in (\ref{egamma}) cancel
each other. This surprising cancellation between contributions of
ghost and chiral fields to one-loop effective action in ${\cal N}=4$
SYM theory was firstly noted in \cite{8} in harmonic superspace
approach. It should be especially pointed out that this result is
correct only on the constant chiral superfield background.

Finally, due to the cancellation between (\ref{fpresult}) and
(\ref{hresult}), the whole one-loop contribution to the effective action
(\ref{egamma}) has an extremely simple form and is determined only by vector loop contribution
\begin{equation}\label{O}
\Gamma = i\sum_{I<J}\,{\rm Tr}\ln (O_{V}-M)_{IJ}
\end{equation}
and all background superfield dependence is encoded in $M$. For
the operator in the above relation, the powers expansion over
Grassmann derivatives of the gauge field strength of the
functional trace has been already calculated by different ways for
models with one chiral background superfield (see \cite{15, ohr, bkts}
and reference therein). As a result, we transformed a rather
complicated problem with hypermultiplet background to a known
problem. The feature of the theory with hypermultiplet background
is the combination (\ref{mass}) ${M}=
(\bar{\Phi}\Phi+\bar{Q}Q+\bar{\tilde{Q}}\tilde{Q})$ which is
invariant under the $R$-symmetry group of ${\cal N}=4$
supersymmetry. That allows application of the results obtained for
${\cal N}=1$ models to the case under consideration making the
corresponding redefinition of the quantity $M$.

The functional trace (\ref{O}) can be written
as a power expansion of dimensionless
combinations $\Psi$, $\bar \Psi$ in hypermultiplet superfields, where
\begin{equation}
{\bar \Psi}^2 = \frac{1}{M^2} \,\nabla^2 {W^2}, \quad \Psi^2 =
\frac{1}{M^2} \,{\bar \nabla}^2 {\bar W}^2.
\end{equation}
In the constant field approximation
this expansion is summed to the following expression for the whole one-loop effective
action (see details in \cite{bkts}):
\begin{equation}\label{n4gamma}
\Gamma = \frac{1}{8\pi^{2}}\int{\rm d}^8 z \int_{0}^{\infty}{\rm
d}t\,t\,{\rm e}^{-t}\frac{ W^{2}\bar{W}^{2}}{M^2}\,\omega (t\Psi,
t\bar{\Psi}),
\end{equation}
$$ \omega (t\Psi, t\bar{\Psi})=
\frac{\cosh(t\Psi)-1}{t^{2}\Psi^{2}}\,\frac{\cosh(t\bar\Psi)
-1}{t^{2}{\bar \Psi}^{2}}\,\frac{t^{2}(\Psi^{2} - {\bar\Psi}^{2})}
{\cosh(t\Psi) - \cosh(t\bar\Psi)}.
$$
As a result, we see that the only difference between the effective actions
with and without the hypermultiplet background is stipulated by the structure of
matrix $M$ defined by (\ref{mass}).
In component form, the
closed relation for one-loop effective action (\ref{n4gamma}) has natural
Schwinger-type expansion over $F^2/M^2$ powers. The expansion
doesn't include $F^6$ term that is a property of ${\cal N}=4$ SYM
theory \cite{bkts, Fr}. The function $\omega$ defined in
(\ref{n4gamma}) (see \cite{bkts}) has the following expansion
\begin{equation}\label{omeg}
\omega (x,y)= \frac{1}{2} + \frac{x^{2}y^{2}}{4\cdot5!} -
\frac{5}{12\cdot7!}\,(x^{4}y^{2} +x^{2}y^{4})
 + \frac{1}{34500\,}(x^{2}y^{6} + x^{6}y^{2}) +
\frac{1}{86400}\,x^{4}y^{4} + \ldots
\end{equation}
Eq. (\ref{omeg}) allows to expand the effective action
(\ref{n4gamma}) in series in powers of $\Psi^2$, $\bar{\Psi}^2$ as
follows
\begin{equation}\label{gdecom}
\Gamma = \Gamma_{(0)}+\Gamma_{(2)}+\Gamma_{(3)}+\cdots,
\end{equation}
where the term $\Gamma_{(n)}$ contains terms
$c_{m,l}\Psi^{2m}\bar{\Psi}^{2l}$ with $m + l= n$. In the bosonic sector,
this expansion corresponds to expansion in powers of the strength $F$,
namely $\Gamma_{(n)}\sim F^{4+2n}/M^{2+2n},\quad
M=(\Phi\bar\Phi+f^{ia}f_{ia})$, where $\Phi, \bar\Phi$ and
$f^{ia}$ are physical bosonic fields of the ${\cal N}=2$ vector
multiplet and hypermultiplet.

\section{Transformation of the ${\cal N}=1$ supersymmetric effective
action to a manifest ${\cal N}=2$ supersymmetric form}
The effective
action (\ref{n4gamma}) and its expansion (\ref{gdecom}) are given
in terms of ${\cal N}=1$ superfields. Our next purpose is to find
a manifest ${\cal N}=2$ form of each term in expansion
(\ref{gdecom}). To do that, we extract from $M$ the ${\cal N}=1$
form of the
$X=-\frac{\bar{Q}Q+\bar{\tilde{Q}}\tilde{Q}}{\bar{\Phi}\Phi}$,
(which was defined in eq. (\ref{2})  in terms of ${\cal N}=2$
superfields) writing $M=\Phi \bar\Phi (1- X)$, and then expand the
denominator $(1/M)^{k}$ from (\ref{n4gamma}) in a power series in
$X$. This expansion leads to the following form for a generic term
of the series
\begin{equation}\label{genterm}
\int d^{8}z\,
\frac{W^2 \bar{W}^2}
{(\Phi\bar{\Phi})^{2(m+l+k+1)}}
(\nabla^2 W^2)^{m}
({\bar{\nabla}}^{2}\bar{W}^2)^l
\left(-(\bar{Q}Q+\bar{\tilde{Q}}\tilde{Q})\right)^{k}
\end{equation}
Further, using $\int{\rm d}^{12} z =\int{\rm d}^8 z
({\nabla}_{2})^{2}({\bar{\nabla}}_{2})^{2}$ and definitions of
${\cal N}=1$ projections for ${\cal N}=2$ on-shell vector
multiplet ${\cal W}|= \Phi, {\nabla}_{2\alpha}{\cal
W}|=-W_{\alpha}, {\nabla}_{2}^{2}{\cal W}|=0$, we can reconstruct
${\cal N}=2$ form of the above generic term. It is worth pointing out that
the reconstruction procedure has some off-shell ambiguity (see
\cite{kth}) even for vanishing hypermultiplet fields but this ambiguity
is unessential in the case under consideration.

The derivative expansion (\ref{n4gamma}) of the effective action
contains the known non-holomorphic potential as a first term (see
(\ref{g0})). It can be unambiguously rewritten in a ${\cal N}=2$
form, following from ${\cal N}=1$ calculations on the
background (\ref{3}). This unique term is automatically ${\cal
N}=4$ supersymmetric since it does not contain the derivatives of the
hypermultiplet and vector strengths. Recovering of the other terms
in the derivative expansion of the effective action is not so
evident and needs special prescriptions.

Calculation of the above effective action was fulfilled on the constant
background (\ref{3}), but for recovering the ${\cal N}=2$ form
such a background is insufficient. We must take into account the
derivatives of the ${\cal N}=1$ hypermultiplet fields. The
procedure of restoring the ${\cal N}=2$ supersymmetric expressions,
based on corresponding ${\cal N}=1$ reduction, always implies
forming of the ${\cal N}=2$ integral measure $\int{\rm d}^{12} z
=\int{\rm d}^8 z ({\nabla}_{2})^{2}({\bar{\nabla}}_{2})^{2}$.
Therefore, to get an integral over ${\cal N}=2$ superspace from an
integral over ${\cal N}=1$ superspace, we must form the derivatives
$({\nabla}_{2})^{2}({\bar{\nabla}}_{2})^{2}$ in the initial ${\cal
N}=1$ superspace integrand. In order to obtain such total derivatives
in the integrand (\ref{genterm}), we have to add all necessary
$\nabla_{\alpha j}\,q^{ia}$ derivative-containing terms with
specified numerical coefficients to the initial ${\cal N}=1$ superspace integrand by hand,
since they did not appear in the process of computations. If we
calculate the effective action in terms of ${\cal N}=1$ superfields not
on the special background (\ref{3}) but on the proper background
(\ref{corset}), these absenting terms will be presented
automatically. Then the derivatives $\nabla_2^2{\bar \nabla}_2^2$
could be formed in the ${\cal N}=1$ superspace integrand and, as a
result, we would obtain the integral over ${\cal N}=2$ superspace.

Further, we use the evident enough assumptions about the properties of
the effective action. Effective action is manifestly ${\cal N}=2$
supersymmetric and, hence, each term in its expansion in
derivatives can be written as the integral over ${\cal N}=2$
superspace of function depending on ${\cal N}=2$ superfield
strengths, hypermultiplet superfields and their spinor
derivatives. It allows to argue as follows. Using
integrations by parts in the integrals over ${\cal N}=2$ superspace and
expressing the terms in derivative expansion of effective action, we
transfer all derivatives from hypermultiplets to the ${\cal N}=2$
superfield strengths and then ones make the reduction to ${\cal N}=1$
form. As a result, we see that all terms in derivative expansion can
be written in the form similar to $\Gamma_{(n)}$ defined in
(\ref{gdecom}), i.e.  without derivatives of the hypermultiplet
superfields. It means that we can act in reverse order beginning with
the given ${\cal N}=1$ form and restoring the corresponding ${\cal N}=2$
form. Also we take into account that the derivative expansion at
vanishing hypermultiplet superfields is presented in terms of ${\cal N}=2$
superconformal scalars \cite{bkts}
\begin{equation} \label{Psi}
{\bf\bar\Psi^2} =\frac{1}{\bar{\cal W}^{2}}\nabla^{4}\ln{\cal W},\:
{\bf\Psi^2} =\frac{1}{ {\cal W}^{2}}\bar\nabla^{4}\ln\bar{\cal W}
\end{equation}
and will search for hypermultiplet dependence compatible with this
property.

Further we demonstrate how the use of the above prescription allows to
obtain the functionals ${\Gamma}_{(0)}, {\Gamma}_{(2)},
{\Gamma}_{(3)}, \ldots$ (\ref{gdecom}) in terms of ${\cal N}=2$
superfields. Let us begin with the functional ${\Gamma}_{(0)} =
\frac{1}{(4\pi)^2}\int d^8z \frac{{W}^2\bar{W}^2}{M^2}$ (which is
$\sim F^4$) and rewrite it in the form (\ref{genterm}) using
$\frac{1}{(1-X)^2} = \sum_{k=0}^{\infty}(k+1)X^k$
\begin{equation}\label{integr}
\frac{1}{(4\pi)^{2}}\int d^8z(
\frac{W^2\bar{W}^2}{\Phi^2\bar{\Phi}^2} +
\sum_{k=1}^{\infty}(k+1)\cdot\frac{W^2\bar{W}^2}{\Phi^{2+k}\bar{\Phi}^{2+k}}
\cdot(-(\bar{Q}Q+\bar{\tilde{Q}}\tilde{Q}))^{k}).
\end{equation}
It is natural to identify the quadratic combination of ${\cal
N}=1$ superfields $(\bar{Q}Q+\bar{\tilde{Q}}\tilde{Q})$ of ${\cal
N}=1$ superfields with ${\cal N}=1$ projection of the quadratic
combination of Fayet-Sohnius hypermultiplets $q^{ia}q_{ia}$.
Such an identification can be checked e.g. by comparison of the
component structures. Then we apply the relations
\begin{eqnarray} \nabla_2^{2}\ln {\cal W}| &=&
-\left({W^{\alpha}W_{\alpha}\over \Phi^{2}}\right)+\ldots,\nonumber\\
\nabla^{2}_{2}{1\over {\cal W}^{m}}| &=& {m
(m+1)\over\Phi^{m}}{W^{\alpha}W_{\alpha}\over
\Phi^{2}}+\ldots,\label{recon}
\end{eqnarray}
where dots mean the terms involving the derivatives of $\Phi$
which can be omitted in our on-shell analysis. Thus, the ${\cal
N}=1$ integrand (\ref{integr}) can be written via ${\cal N}=2$
vector multiplet superfields and hypermultiplets as
\begin{eqnarray}
\nabla_2^2\ln {\cal W}\, \bar{\nabla}_2^2 \ln
\bar{\cal W} + \sum_{k=1}^{\infty} \frac{1}{k^2(k+1)}\nabla_2^2
\frac{1}{{\cal W}^k}\bar{\nabla}_2^2 \frac{1}{\bar{\cal W}^k}
\cdot(-q^{ia}q_{ia}) + ...,
\end{eqnarray}
where the dots mean all terms involving hypermultiplet derivatives
of the form $$\nabla_2^{\alpha} \frac{1}{{\cal W}^k} \nabla_{2
\alpha} (-q^{ia}q_{ia}) \bar{\nabla}_2^2 \frac{1}{\bar{\cal
W}^k},\; \frac{1}{{\cal W}^k} \nabla_2^2 (-q^{ia}q_{ia})
\bar{\nabla}_2^2  \frac{1}{\bar{\cal W}^k},$$ which, according to
the above prescriptions, should be added in order to obtain the
full  ${\cal N}=2$ integration measure $\nabla^2_2\bar\nabla^2_2$
in the integral over ${\cal N}=1$ superspace (40). As a result,
the above prescriptions lead to an expression
\begin{equation}\label{g0}
\Gamma_{(0)}={1\over (4\pi)^{2}}\int d^{12}z (\ln {\cal W}\ln
\bar{\cal W}+ \sum_{k=1}^{\infty}{1\over k^{2}(k+1)} X^{k}),
\end{equation}
where $X=\left(-{q^{ia}q_{ia}\over {\cal W}\bar{\cal
W}}\right)$ was defined in (\ref{2}). The second term in
(\ref{g0}) can be transformed to the form (\ref{2}) using the power
series for the Euler dilogarithm function and relation
$\frac{1}{k^2(k+1)}=\frac{1}{k^2}-\frac{1}{k}+\frac{1}{k+1}$. We see
that the expression (\ref{g0}) is just the effective Lagrangian
(\ref{2}) found in \cite{33}.

${\cal N}=2$ form of next term ($\sim F^8$) in the series
(\ref{gdecom}) is reconstructed using (\ref{recon}) and expansion of
$(1/M)^{6}$ in $X$. Direct analysis, analogous to one
in the previous case, leads to the following expression for
$\Gamma_{(2)}$ in (\ref{gdecom})
\begin{equation}\label{g2}
\Gamma_{(2)}= {1\over 2(4\pi)^{2}}\int d^{12}z
{\bf\Psi^{2}\bar{\Psi}^{2}} ({1\over 36}  + {1\over
5!}\sum_{k=1}^{\infty}{(k+5)(k+4)(k+1)\over (k+3)(k+2)}X^{k}).
\end{equation}
The
$X$-independent part of this term was given in \cite{bkts}.  The sum in
(\ref{g2}) can be written in an explicit form as follows
\begin{equation}\label{sum} \sum_{k=1}^{\infty}{(k+5)(k+4)(k+1)\over (k+3)(k+2)}X^{k}= $$ $$
={1\over (1-X)^{2}}+{4\over (1-X)}+ {6X-4\over X^{3}}\ln
(1-X)+4{X-1\over X^{2}}-{10\over 3}.
\end{equation}
Applying
the same procedure to the third term ($\sim F^{10}$) in (\ref{gdecom}),
one obtains \begin{equation} \Gamma_{(3)}=-{5\over 6\,(4\pi)^{2}}\int
d^{12}z ({\bf\Psi^{4}\bar{\Psi}^{2}+\Psi^{2}\bar{\Psi}^{4}})(-{1\over
5!} + {1\over 7 !}\sum_{k=1}^{\infty}(k+7)(k+6)(k+1)X^{k}),
\end{equation}
where the sum in the right-hand side is
\begin{equation}
\sum_{k=1}^{\infty}(k+7)(k+6)(k+1)X^{k} ={2X\over
(1-X)^{4}}(56-116X+84X^{2}-21X^{3}).
\end{equation}
Thus, we have found the hypermultiplet-dependent complementary terms to
$\Gamma_{(0)}$, $\Gamma_{(2)}$ and $\Gamma_{(3)}$ for the effective
action obtained in \cite{bkts} in the ${\cal N}=2$ vector multiplet sector.
Clearly every term in the expansion of effective action
(\ref{gdecom}) can be written in ${\cal N}=2$ supersymmetric form.
For example, the $X$-dependent part of the fourth term ($\sim
F^{12}$) in (\ref{gdecom}) contains two parts. The first one is
\begin{equation}\label{g4-1}
\Gamma_{(4_{1})}={1\over (4\pi)^{2}}{1\over
17250}\int d^{12}z ({\bf
\Psi^{2}\bar{\Psi}^{6}+\Psi^{6}\bar{\Psi}^{2}})\times
\end{equation}
$$
\times{12X\over(1-X)^{6}} (450-1545X+2284X^{2}-
1779X^{3}+720X^{4}-120X^{5})
$$
and the second part is given as follows
\begin{equation}\label{g4-2}
\Gamma_{(4_{2})}={1\over 5\cdot 6!}{1\over
(4\pi)^{2}}\int d^{12}z {\bf \Psi^{4}\bar{\Psi}^{4}}\times
\end{equation}
$$
\times( {12(5X-4)\over X^{5}}\ln (1-X)- {1\over 5
X^{4}(1-X)^{6}}(240-1620X+4610X^{2}-
$$
$$
-7120X^{3}+6363X^{4}-4878X^{5}+ 6135X^{6}-
7560X^{7}+5670X^{8}-2268X^{9}+378X^{10})).
$$

Thus, we see that this ${\cal N}=2$ reconstruction procedure for
the effective action (\ref{n4gamma}) of ${\cal N}=4$ SYM theory,
written initially in terms of ${\cal N}=1$ superfield, can be
realized completely for any terms in the expansion (\ref{gdecom})
completing these terms by the corresponding terms containing the
hypermultiplet superfields.  Unfortunately, we can not guarantee
that the reconstructed effective action will be ${\cal N}=4$
invariant. Therefore this point needs independent test. For this
purposes, we can use either ${\cal N}=1$ form of the hidden ${\cal
N}=4$ supersymmetry transformations (\ref{hidden},
\ref{manifest}), or ${\cal N}=2$ form of hidden ${\cal N}=2$
supersymmetry transformations (\ref{hidden2}) in the harmonic
superspace \cite{33}.

The low-energy effective action ${\cal N}=4$ SYM theory is
expected to be self-dual and, in addition, invariant under the
(probably quantum deformed, see \cite{KuzMcTh}) superconformal group transformations.
But it turns out that even these requirements are not sufficient
to fix the ${\cal N}=4$ form of the effective action functional
uniquely \cite{kth, kul}. Until now we have used the constant
field approximation (\ref{3}) which supposes that all derivatives
of the hypermultiplet fields vanish. This approximation suffices
to restore the manifestly ${\cal N}=2$ supersymmetric effective
action on the basis of its ${\cal N}=1$ form (\ref{n4gamma}),
calculated on the background (\ref{3}), in terms of ${\cal N}=2$
superconformal scalars \cite{bkts}. However, finding ${\cal N}=4$
supersymmetric effective action requires fulfillment of all
calculations on the background mentioned in the section 2.2. (i.e.
supposing $Dq \neq 0$). Another way to obtain terms with
derivatives $Dq^{ia}$, necessary for constructing the ${\cal N}=4$
supersymmetric form, can be based on algebraic consideration
analogous to \cite{31}.

Since in on-shell description the  hypermultiplet superfields
$q_{i\,a}$ and superfield strengths ${\cal W}, \bar{\cal W}$ are
independent on harmonic variables $u^\pm_i$, one can insert a
harmonic integral into the expressions for $\Gamma_{(0)},\,
\Gamma_{(2)},\, \Gamma_{(3)}, \ldots$ and write the variables $X$
as  $X = \left(-2{q^{+a}{q}^{-}_{a}\over {\cal W}\bar{\cal
W}}\right)$. This allows studying the variation of effective
action under hidden supersymmetry transformations using harmonic
superspace formalism. This variational procedure and other problems
of restoring ${\cal N}=4$ supersymmetric effective action will be
considered in the next section.

\section{Problem of restoring ${\cal N}=4$ supersymmetric effective
action}
As we already pointed out, all manifestly ${\cal N}=2$
supersymmetric contributions obtained in the previous section and
defining a derivative expansion, except (\ref{g0}), should not be ${\cal
N}=4$ supersymmetric because of the background choice (\ref{3}) and
the gauge-fixing procedure (\ref{f}). In particular, they must be
non-invariant under hidden ${\cal N}=2$ supersymmetry transformations
(\ref{hidden2}), except (\ref{g0}). It is obvious that in order to
obtain ${\cal N}=4$ supersymmetric contributions from the ones given in the
previous section, we have to add to each term in the derivative
expansion of (\ref{n4gamma}) some extra terms.
On equal footing these extra terms must contain fields $\lambda = W|$ of the vector
multiplet, which are presented in the effective action (\ref{n4gamma}),
as well as fields $\psi = D q|$ of the hypermultiplet, which are absent in the
special background definition (\ref{3}).

We study a possible form of terms depending on the
hypermultiplet $q^{+}$ and its spinor derivatives $D^{-}q^{+}$ in
the effective action. We assume that this on-shell ${\cal N}=4$
supersymmetric effective action is described by manifestly ${\cal
N}=2$ supersymmetric effective Lagrangian depending on ${\cal W},
\bar{\cal W}$, their spinor derivatives, $q^{+}$ and spinor
derivatives of $q^{+}$. The leading low-energy contribution to
such an effective Lagrangian is known and given by the expressions
(\ref{1}), (\ref{2}). Now we discuss a possibility to obtain
next-to-leading corrections. First, we note that the $D^{-}q^{+}$
is a fermionic superfield; second, it is known that on shell
$(D^{-})^{2}q^{+}=0$ \cite{33}. Hence, the effective Lagrangian
can be written as a finite order polynomial in first derivatives
$D^{-}q^{+}$ with the coefficients depending on $q^{+}$ and ${\cal
W}, \bar{\cal W}$ and their spinor derivatives. Since the
superfield effective Lagrangian is integrated over full ${\cal
N}=2$ superspace, it is dimensionless and chargeless. The only way
to compensate dimensional quantities $D^{-}q^{+},
(D^{-}q^{+})^{2}, ...$ is to use the ${\cal N}=2$ strengths and
their spinor derivatives. Therefore we can write down all the
possible terms in expansion of the effective Lagrangian in power
series in $D^{-}q^{+}$ up to dimensionless functions which can
depend only on dimensionless quantities $X$ (\ref{2}) and
quantities (\ref{Psi}). As a result, we get a method allowing, in
principle, to find the entire structure of the on-shell ${\cal N}=4$
supersymmetric effective action.

Let's assume that the above procedure has been realized and we got
such an effective Lagrangian. Using the integrations by path we
transfer, where it is possible, all spinor derivatives from
$q^{+}$ onto ${\cal W}, \bar{\cal W}$ and compare a result with
derivative expansion of the expression (\ref{n4gamma}). Since the
effective Lagrangian under consideration contains the derivatives
$D^{-}q^{+}$ which are absent in (\ref{n4gamma}), transferring the
derivatives from $q^{+}$ to ${\cal W}, \bar{\cal W}$ will lead to
new terms in comparison to (\ref{n4gamma}). Hence, one can expect
that a proper ${\cal N}=4$ supersymmetric effective Lagrangian, in
principle, should have the other numerical coefficients at the
$X$-dependent terms in comparison to (\ref{n4gamma}), except the
term (\ref{2}). We conclude once more that the result
(\ref{n4gamma}) does not allow to derive the ${\cal N}=4$ supersymmetric
effective action directly.

However, there exists a principal possibility to construct
${\cal N}=4$ supersymmetric on-shell effective action on the basis
of the relation (\ref{n4gamma}). We can act as follows. Let us
consider a derivative expansion of (\ref{n4gamma}) at $X=0$.
Each term of the expansion is expressed in a manifestly ${\cal
N}=2$ supersymmetric form \cite{bkts}. We insert the integral over
harmonics, which is equal to unit, into each integral over full
${\cal N}=2$ superspace, taking into account that the integrands
are harmonic-independent. Then we investigate which
hypermultiplet-dependent terms have to be added to each term
of the derivative expansion so that the whole expansion would become
invariant under the hidden ${\cal N}=2$ supersymmetry transformations.
This is just the procedure proposed in \cite{31} for finding
the ${\cal N}=4$ extension of non-holomorphic effective potential
(\ref{1}). Further, we are going to discuss an application of this procedure to
reconstruction of some leading contributions to the on-shell ${\cal
N}=4$ extension of $F^{8}$-term in effective action.

Since the entire effective action can be presented as a polynomial
in $D^{-}q^{+}$, terms in the effective action corresponding
to the ${\cal N}=4$ extension of $F^{8}$ have a form
\begin{equation}\label{49}
g(X){\bf\Psi^2}{\bf\bar\Psi^2} + ...,
\end{equation}
where the dots mean the terms depending on the derivatives
$D^{-}q^{+}$. Direct quantum field calculation of the
$D^{-}q^{+}$-dependent terms demands to use an appropriate background
but it is an unsolved problem yet.\footnote{To get such an effective
action in ${\cal N}=1$ formalism we have to carry out the calculations
keeping the spinor derivatives of background chiral superfields. The
only example of these calculations was given within Wess-Zumino model
for finding the effective potential of auxiliary fields in Refs.
\cite{bky, 15}. In particular, such a potential for chiral ${\cal N}=1$
superfields of the ${\cal N}=2$ vector multiplet arises from the
self-dual requirement for the ${\cal N}=4$ SYM effective action (see
Ref. \cite{kul}).}
Here we consider calculation of the first,
derivative-independent term only $g(X){\bf\Psi^2}{\bf\bar\Psi^2}$ in
(\ref{49}) and show that the function $g(X)$ can be reconstructed on
the basis of $X$-independent term in (\ref{g2}). Then we check if
this function $g(X)$ coincides with $X$-dependent terms in (\ref{g2}).
Taking into account that we are interested only in the first term in
the derivative expansion (\ref{49}), we systematically omit, in process
of our calculations, all terms containing $D^{-}q^{+}$, since they
can not contribute to $g(X)$.

Let us consider the variation of the $X$-independent term $\sim{\bf
\Psi}^{2}{\bf \bar\Psi}^{2}$ in (\ref{g2}) under hidden
${\cal N}=2$ transformation with parameters $\varepsilon^{\alpha
a}$ directly in ${\cal N}=2$ harmonic superspace. Writing $\bar{D}^4$ as
$\frac{1}{4}\bar{D}^{+2}\bar{D}^{-2}$ allows to rewrite the
initial expression as $ \bar{D}^{+2}\frac{1}{\bar{\cal
W}^2}\bar{D}^{-2}\ln (\bar{\cal W}), $ which is equal to
$$
-2\cdot 3 \frac{\bar{D}^{+ \dot\alpha}\bar{\cal W}
\bar{D}^{+}_{\dot\alpha}\bar{\cal W}}{\bar{\cal W}^4}\cdot
\frac{\bar{D}^{- \dot\beta}\bar{\cal W}
\bar{D}^{-}_{\dot\beta}\bar{\cal W}}{\bar{\cal W}^2}.
$$
Because of $\delta \bar{\cal W} \sim D ^{-}_{\alpha}q^+$, a variation
of  $\bar{D}\bar{\cal W}$ is proportional to $\sim D_{\alpha
\dot\alpha}q^+$. Such terms are systematically truncated and
the whole variation is defined only by the numerator variation, which
gives
$$
2\cdot3\cdot6 \frac{\delta \bar{\cal W}}{\bar{\cal
W}^7}\bar{D}^+ \bar{\cal W}\bar{D}^+ \bar{\cal W}\bar{D}^-
\bar{\cal W}\bar{D}^- \bar{\cal W}.
$$
After integration by parts, the last expression becomes
$$
-3\frac{\delta \bar{\cal
W}}{\bar{\cal W}^3}\bar{D}^4 \ln \bar{\cal W }.
$$
Collecting all variations, we get
\begin{equation}\label{first}
\delta\frac{1}{36}{\bf\Psi}^2{\bf\bar{\Psi}}^2 = (\frac{1}{12}q^{+
a}\varepsilon^{\alpha}_a D^-_{\alpha}{\cal W})\frac{D^4 \ln{\cal
W}}{{\cal W}^3}\frac {\bar{D}^4\ln \bar{\cal W}}{\bar{\cal W}^3}.
\end{equation}
One can note that the hypermultiplet fields variation in the first
$X$-dependent term $ {1\over 24}{\bf \Psi}^{2}{\bf
\bar\Psi}^{2}\left({-2q^{a +}q^{-}_{a}\over {\cal W}\bar{\cal
W}}\right)$ in (\ref{g2}) gives the expression similar to
(\ref{first}) but with another factor: $-1/24$. To obtain this
result, we used a property of the full harmonic superspace integral
$\int d u \;\delta (q^{+ a} q^-_a) = \int d u \;(\delta q^{+ a}
D^{--}q^+_a + q^{+ a} \delta q^-_a) = \int d u\; 2 q^{+ a} \delta
q^-_a$. It shows that the variation of the first term in (\ref{g2})
does not cancel the variation of the linear in $X$ terms in the
$X$-dependent sum. It means that the invariance under above
transformations is impossible. It is felt that this
non-invariance is compensated by "quantum" modification
$\delta_{mod} = \delta + \delta_{q}$ of the classical
transformation law like $$ \delta_{q}{\cal W} = c\frac{1}{
\bar{\cal W}^2}D^4\frac{1}{{\cal W}^2} (\bar{D}^4 \ln \bar{\cal
W})(\epsilon^{\dot\alpha a}D^{-}_{\dot\alpha} q^{+}_{a}), $$ and
these quantum additions to the variation of $\Gamma_{(0)}$ would
compensate the non-invariance of the variation $\Gamma_{(2)}$
(\ref{g2}). However, the direct analysis of several first terms in sums (\ref{g0}) and (\ref{g2}) shows that the quantum modification
can not save the situation since a coefficient $c$ presenting in
$\delta_q$ is changed with every order of $X$ and never can be
chosen properly. We see the ${\cal N}=4$ invariant expressions
can't be constructed by the naive quantum modification of the
classical transformations (\ref{hidden}, \ref{hidden2}) and
therefore the quantum modification problem should be studied
separately. Nevertheless, the algebraic procedure proposed in
\cite{31} for ${\cal N}=4$ completion can be applied in the case
under consideration.

To compensate the expression in right hand side of (\ref{first}),
we introduce the $X$-dependent
complimentary term
\begin{equation}\label{i1}
I_{1}=C_{1}{\bf \Psi}^{2}{\bf \bar\Psi}^{2}\left({-2q^{a
+}q^{-}_{a}\over {\cal W}\bar{\cal W}}\right).
\end{equation}
We demand, a variation of the hypermultiplet "numerator" (i.e.
$-2q^{a +}q^{-}_{a}$) in (\ref{i1}) should compensate the
variation (\ref{first}). This requirement fixes the coefficient
$C_{1}=1/12$. But the whole variation $I_{1}$ again contains another
extra variation terms, which can be compensated only by
introducing one more complementary terms. Further, we
consider a  variation of $\bar{\cal W}$ in (\ref{i1}). The
expression
$$ \frac{q^+q^-}{\bar{\cal W}^3}\bar{D}^4 \ln\bar{\cal W} =
-{1\over 2}\bar{D}^{+2}(\frac{q^+q^-}{\bar{\cal
W}^3})\frac{\bar{D}^-\bar{\cal W}\bar{D}^-\bar{\cal W}}{\bar{\cal
W}^2}
$$
can be written as
\begin{equation}\label{comp_part}
{1 \over 4}\left(
6q^{+}\bar{D}^{+\dot{\alpha}}q^{-} {\bar{D}^{+}_{\dot\alpha}\bar{\cal W}\over \bar{\cal W}^6}
- 12{q^{+}q^{-}\over \bar{\cal W}^7}\bar{D}^{+ \dot\alpha}\bar{\cal W}
\bar{D}^{+}_{\dot\alpha}\bar{\cal W}
\right)\bar{D}^{-\dot\beta}\bar{\cal W}\bar{D}^{-}_{\dot\beta}\bar{\cal
W}.
\end{equation}
The numerator variations of both terms in (\ref{comp_part})
give
$$
{-2 \over 12\cdot 4}\cdot {D^4 \ln {\cal W}\over {\cal W}^3}\left(
6\cdot 6 \;\delta \bar{\cal W} q^{+}\bar{D}^{+\dot{\alpha}}q^{-} {\bar{D}^{+}_{\dot\alpha}\bar{\cal W}\over \bar{\cal W}^7}
+ 12\cdot 7\; \delta \bar{\cal W}{q^{+}q^{-}\over \bar{\cal W}^8}\bar{D}^{+ \dot\alpha}\bar{\cal W}
\bar{D}^{+}_{\dot\alpha}\bar{\cal W}
\right)\bar{D}^{-\dot\beta}\bar{\cal W}\bar{D}^{-}_{\dot\beta}\bar{\cal
W}.
$$
The first term in the brackets is represented by means of the
superconformal invariants (\ref{49})
\begin{equation}\label{f_term}
-{3 \over
4}\cdot { \delta \bar{\cal W}\over \bar{\cal W}^4}\cdot {D^4 \ln {\cal
W}\over {\cal W}^3} q^{+}\bar{D}^{+\dot\alpha}q^{-}\cdot
\bar{D}^{+}_{\dot\alpha}\bar{D}^{-2}\ln \bar{\cal W}
\end{equation}
as well as the second term
\begin{equation}\label{s_term}
{7 \over 2\cdot 5}\cdot { \delta \bar{\cal W}\over \bar{\cal
W}^4}\cdot {D^4 \ln {\cal W}\over {\cal W}^3} \left(
q^{+}\bar{D}^{+\dot\alpha}q^{-}\cdot
\bar{D}^{+}_{\dot\alpha}\bar{D}^{-2}\ln \bar{\cal W} +
(q^{+}q^{-})\bar{D}^4 \ln {\cal W} \right).
\end{equation}
In order to compensate
\begin{equation}\label{extra_term}
{7 \over 4\cdot 5}\cdot {\varepsilon^{\alpha
a}D^{-}_{\alpha}q^{+}_{a} (q^{+}q^{-})\over {\cal W}^3} {D^4 \ln
{\cal W}\bar{D}^4 \ln \bar{\cal W}\over \bar{\cal W}^4},
\end{equation}
which is a part of (\ref{s_term}), we add the next complementary term
\begin{equation}\label{complement_2}
I_{(2)}= c_{2}\left({-2(q^{+}q^{-})\over {\cal W}\bar{\cal
W}}\right)^{2} {1\over {\cal W}^2}\bar{D}^4 \ln \bar{\cal W} {1\over \bar{\cal W}^2}
D^4 \ln {\cal W},
\end{equation}
Then we consider variation of $(q^+q^-)$ in this term and compare a
result to (\ref{extra_term}). The requirement of compensation
fixes the coefficient $C_{2}= {7\over 2\cdot 4\cdot 5}$. The
uncompensated part in variation of $I_{(2)}$ is
\begin{equation}\label{extra_term_2}
-{1\over 2\cdot 4\cdot 5}\cdot
\varepsilon^{\alpha a}
\left(
{3 q^{+}_{a}D^{-}_{\alpha}{\cal W}\over {\cal
W}^4}q^{+}\bar{D}^{+\dot\alpha}q^{-}
- {q^{+}_{a}\over {\cal W}^3}(D^{-}_{\alpha}q^{+}\bar{D}^{+\dot\alpha}q^{-})
\right)\bar{D}^{+}_{\dot\alpha}\bar{D}^{-2} \ln \bar{\cal W}.
\end{equation}
In order to compensate the first term in the last variation, we
introduce another type complementary term
\begin{equation}\label{complement_d1}
J_{(1)}^{0}= d_1 \left[\left({-2q^{+}q^{-}\over {\cal W}^4 \bar{\cal W}^4}\right)
(q^{+}\bar{D}^{+\dot\alpha}q^{-})\right]\bar{D}^{+}_{\dot\alpha}\bar{D}^{-2}
\ln \bar{\cal W} D^4 \ln {\cal W}.
\end{equation}
In the component form the term $J_{(1)}^{0}$ is proportional to
$$
{F^{4}\bar{F}^{2}\over \Phi^8 \bar\Phi^8}(f\bar{f})
f^{a}\bar\kappa_{a}^{\dot\alpha}\bar{F}^{\dot\beta}_{\dot\alpha}\bar{\lambda}_{\dot\beta},
$$
which obviously vanishes in the bosonic sector of the theory.
A partial variation obtained by varying only $q^+$ in square
brackets (\ref{complement_d1}) leads to $$ \tilde\delta
J_{(1)}^{0} = d_1 \left({q^{+}_{a}\varepsilon^{\alpha a}
D^{-}_{\alpha}{\cal W} \over {\cal W}^4 \bar{\cal W}^4}\right)
(q^{+}\bar{D}^{+\dot\alpha}q^{-})\bar{D}^{+}_{\dot\alpha}\bar{D}^{-2}
\ln \bar{\cal W} D^4 \ln {\cal W} + $$
\begin{equation}\label{j_part2}
+ d_1 \left({-2q^{+}q^{-}\over {\cal W}^4 \bar{\cal W}^4}\right)
\left( {1\over 4}\varepsilon^{\alpha a} D^{+}_{\alpha}{\cal
W}\bar{D}^{+\dot\alpha}q^{-}_{a}\bar{D}^{+}_{\dot\alpha}\bar{D}^{-2}
\ln \bar{\cal W} D^4 \ln {\cal W}. \right)
\end{equation}
In order to compensate the first term in (\ref{extra_term_2}), the
coefficient $d_1 = {3\over 2\cdot 4\cdot 5}$. The direct
calculation shows that this coefficient allows to compensate the
second term in (\ref{extra_term_2}) with (\ref{j_part2}) as well.
Further, we have to obtain the whole variation of $J_{(1)}^{0}$
and take into account variation of ${\cal W}$ contained in
$J_{(1)}^{0}$. Obviously, in order to compensate the whole variation of
$J^{0}_{(1)}$, the
new complementary terms $J_{(1)}^{n}\sim (Dq)^n$ are necessary. They
will lead to a new type of complementary terms $J_{(n)}^{k}\sim
(Dq)^{n}X^{k}$, where $n=1,2,\dots 8$, $k=0\ldots \infty$.

The analysis of all complimentary terms $I_n$ is greatly
simplified when we calculate only the first $D^+q^{-}$-independent
term in the expansion (\ref{49}). In this case, we can find all
$I_{n}$ in the series of the complementary terms
\begin{equation}\label{complem_1}
I=\sum_{n=0}^{\infty}I_{n}=\sum_{n=0}^{\infty}C_{n}{\bf
\Psi}^{2}{\bf \bar\Psi}^{2}\left({-2q^{a +}q^{-}_{a}\over {\cal
W}\bar{\cal W}}\right)^{n}.
\end{equation}
Let us consider the variation of general term in this series
\begin{equation}\label{rest_gen}
\delta I_{n}= \delta_{1} + \delta_{2}=
I_{n}\left[-{(n+2)(n+6)\over(n+4)}{\delta \bar{\cal W}\over
\bar{\cal W}}\right]  + C_{n}{\bf \Psi}^{2}{\bf
\bar\Psi}^{2}\left({-2q^{b +}q^{-}_{b}\over {\cal W}\bar{\cal
W}}\right)^{n-1}[-4n {q^{a +}\delta q^{-}_{a}\over {\cal
W}\bar{\cal W}}].
\end{equation}
Using the transformations (\ref{hidden2}), we rewrite the second term in
(\ref{rest_gen}) as follows
\begin{equation}\label{gen_rest1}
\delta_{2}=C_{n}{\bf \Psi}^{2}{\bf \bar\Psi}^{2}\left({-2q^{b
+}q^{-}_{b}\over {\cal W}\bar{\cal W}}\right)^{n-1}[-n {q^{a
+}\varepsilon^{\alpha}_{a}D^{-}_{\alpha}{\cal W}\over {\cal
W}\bar{\cal W}}].
\end{equation}
The variation $\delta \bar{\cal W}$ contained in the first term of
(\ref{rest_gen}) is proportional to $D_{\alpha}q$. Let us
transform the first term to form (\ref{gen_rest1}) using
integration by parts. Excepting the superfields ${\cal W}$ and
derivatives $D^4, {\bar D}^4$, which are unessential for this
transformation, we rewrite the first term in (\ref{rest_gen}) in the
form
\begin{equation}\label{ident} \delta_{1}\sim
\frac{(q^{+b}q^-_b)^nD^-_{\alpha}q^+_a}{{\cal
W}^{n+2}}=(n+2)\frac{(q^{+b}q^-_b)^nq^+_a D^-_{\alpha}{\cal W
}}{{\cal W}^{n+3}}-\frac{n(q^{+c}q^-_c)^{n-1}}{{\cal
W}^{n+2}}D^-_{\alpha}q^{+b}q^-_b q^+_a.
\end{equation}
Taking into account the property $D^{--}q^{a+}=q^{a-}$,
$D^-_{\alpha} q^-_a =0$ and cyclicity property $\epsilon_{a
b}\epsilon_{c d} + \epsilon_{c a} \epsilon_{b d}+ \epsilon_{b c}
\epsilon_{a d} =0$ at permutation of  $SU(2)$-isospinor group
indices, we find, after some algebraic manipulations,\footnote{ The
property $D^{--}(q^{+ a}q^-_{a}) =0$ and cyclicity lead to an
identity $D^-_{\alpha}q^{+b}q^-_bq^+_a
-D^-_{\alpha}q^{+b}q^+_bq^-_a -D^-_{\alpha}q^+_aq^{+b}q^-_b=0$.
The second term in this identity can be transformed to
$D^{-}_{\alpha}q^{+ b}q^{-}_{b} q^{+}_{a}$ using integration by
parts and relation $\int du D^{--}(...)=0$. Then, the identity
takes the form
$2D^-_{\alpha}q^{+b}q^-_{b}q^+_a=D^-_{\alpha}q^+_a(q^{+b}q^-_b)$.
Substituting it to the right-hand side of the expression
(\ref{ident}) instead of the second term and transferring a result
to the left-hand side, we obtain
$(1+n/2)\frac{(q^{+b}q^-_b)^nD^-_{\alpha}q^+_a}{{\cal W}^{n+2}}
=(n+2)\frac{(q^{+b}q^-_b)^nq^+_aD^-_{\alpha}{\cal W}}{{\cal
W}^{n+3}}$ that leads to (\ref{gen_rest2}). } the first term of
(\ref{rest_gen}) in the form similar to (\ref{gen_rest1})
\begin{equation}\label{gen_rest2}
\delta_{1}=C_{n}{\bf \Psi}^{2}{\bf \bar\Psi}^{2}\left({-2q^{b
+}q^{-}_{b}\over {\cal W}\bar{\cal
W}}\right)^{n}{(n+2)(n+6)\over(n+4)}{q^{a
+}\varepsilon^{\alpha}_{a}D^{-}_{\alpha}{\cal W}\over{\cal W}
\bar{\cal W}}.
\end{equation}
The requirement of cancellation of (\ref{gen_rest1}) and
(\ref{gen_rest2}) leads to a recursion condition
\begin{equation}\label{rec1}
C_{n}=C_{n-1}{(n+1)(n+5)\over n(n+3)},
\end{equation}
which has a solution
\begin{equation}\label{rec1}
C_{n}={1\over 6\cdot5!}(n+5)(n+4)(n+1).
\end{equation}
It is highly amazing that the correct coefficient (\ref{rec1})
obtained from invariance under hidden ${\cal N}=2$ supersymmetry
transformations differs from the coefficient in (\ref{g2}) only by
the numerical denominator!

Summing the series (\ref{complem_1}}), we find the correct leading
part $\sim g (X)$ in expansion (\ref{49}) of on-shell ${\cal N}=4$
supersymmetric $F^8$-term in the closed form
\begin{equation}\label{59}
I={1\over 72}{1\over (4\pi)^{2}}\int d^{12}z du\,{\bf
\Psi}^{2}{\bf \bar\Psi}^{2}{1-X+{3\over10}X^2\over(1-X)^4}.
\end{equation}
It is obvious that this expression does not coincide with result
(\ref{g2},\ref{sum}) obtained by restoring the ${\cal N}=2$ form
of $\Gamma_{(2)}$ from its ${\cal N}=1$ form. Thus, the leading
bosonic part of complete on-shell ${\cal N}=4$ supersymmetric
extension of $F^8$ invariant is finally established.

\section{Summary}
We have studied the one-loop effective action in ${\cal N}=4$ SYM
theory, depending on ${\cal N}=2$ vector multiplet and hypermultiplet
fields.  The theory under consideration was formulated in ${\cal N}=1$
superspace and quantized in the framework of the background field method with
the use of a special gauge fixing conditions preserving manifest ${\cal N}=1$
supersymmetry. The effective action is given by superfield functional
determinants. The concrete calculations of these determinants are done
on specific ${\cal N}=1$ superfield background corresponding to
constant Abelian strength $F_{mn}$ and constant hypermultiplet fields.
We have proved that the effective action depending on all
fields of ${\cal N}=4$ vector multiplet is restored on the base of
calculations only in ${\cal N}=2$ vector multiplet sector by special
change of functional arguments (see (\ref{O}) and (\ref{n4gamma})).

We have examined a possibility to present the effective action
obtained in a manifest ${\cal N}=2$ supersymmetric form. Analyzing
the effective action as an expansion in spinor covariant
derivatives, we have showed that the terms of this expansion can be
expressed via integrals over ${\cal N}=2$ superspace of the
functions depending on ${\cal N}=2$ strengths, their spinor
derivatives and hypermultiplet superfields. As one of the results,
we have rederived the complete ${\cal N}=4$ supersymmetric
low-energy effective action, which was discovered in \cite{31}.
All other terms in the derivative expansion of the effective
action describe the next-to-leading corrections to the effective
action found.

We point out that all terms in hypermultiplet sector in derivative
expansion of the effective action are gauge-dependent, except the first
leading term. They do not invariant under hidden ${\cal N}=2$
supersymmetry transformations, which are a part of complete
on-shell ${\cal N}=4$ supersymmetry transformations of ${\cal
N}=4$ SYM theory, because of the chosen background and the gauge
fixing procedure. To analyze a possibility of presenting
the derivative expansion terms in on-shell ${\cal N}=4$
supersymmetric form, we applied a formalism of harmonic superspace
and algebraic approach developed in \cite{31}. We have considered the
first subleading term in expansion of the effective action in
${\cal N}=2$ vector multiplet sector ($F^{8}$-term written via
${\cal N}=2$ superconformal invariants depending on strengths
${\cal W}, \bar{\cal W}$ and their spinor derivatives \cite{bkts})
and proved that it can be completed up to on-shell ${\cal N}=4$
supersymmetric form by the hypermultiplet dependent terms and
presented as polynomial in hypermultiplet spinor derivatives.
The first leading term of this polynomial, which depends on
hypermultiplet but does not depend on its derivatives, is given in
explicit form (see (\ref{59})).

The most important extension of this work that may clarify a
structure of next-to-leading corrections to low-energy effective
Lagrangian (\ref{2}) in hypermultiplet sector is a computation of
effective action on the background (\ref{corset}). It can provide
a direct independent verification of the results given in Section
4. It would be extremely interesting also to study full ${\cal
N}=4$ completion of the $F^8$ and higher invariants by all
appropriate hypermultiplet derivative dependent terms.

\section{Acknowledgements}
I.L.B would like to thank E.A. Ivanov, S.M. Kuzenko, A.Yu. Petrov and
A.A. Tseytlin for numerous discussions on the problem of effective
action in extended supersymmetric field theories. The work was
supported in part by INTAS grant, INTAS-00-00254 and RFBR grant,
project No 03-02-16193.  I.L.B is grateful to RFBR grant, project No
02-02-04002 and to DFG grant, project No 436 RUS 113/669 for partial
support. The work of N.G.P and A.T.B was supported in part by RFBR
grant, project No 02-02-17884. I.L.B is grateful to Center of String
and Particle Theory at University of Maryland, where part of this work
has been fulfilled, for partial support and S.J.  Gates for kind
hospitality.  He is also grateful for partial support to INFN,
Laboratori Nazionali di Frascati, where the work was finalized, and S.
Bellucci for warm hospitality.

\end{document}